\documentclass[a4paper,12pt]{article}
\usepackage{bm}
\usepackage{amsmath}
\usepackage{amssymb}
\usepackage[dvips]{graphicx}
\usepackage{pdflscape}
\usepackage{multirow}
\usepackage[sort]{natbib}

\newcommand{\mpr}{\bm{M}^{'}}
\newcommand{\hsp}{\hspace{0.2cm}}
\newcommand{\bx}{\bm{x}}
\newcommand{\by}{\bm{y}}
\newcommand{\bmu}{\bm{\mu}}
\newcommand{\btau}{\bm{\tau}}
\newcommand{\bA}{\bm{A}}
\newcommand{\bepsilon}{\bm{\epsilon}}
\newcommand{\bM}{\bm{M}}

\title{Bayesian Protein Sequence and Structure Alignment}
\author{Christopher J. Fallaize\thanks{%
School of Mathematical Sciences, University of Nottingham, University Park,
Nottingham, NG7 2RD, UK, email: Chris.Fallaize@nottingham.ac.uk} \and Peter J. Green\thanks{%
University of Bristol and University of Technology Sydney} \and Kanti V. Mardia\thanks{%
University of Leeds and University of Oxford} \and Stuart Barber\thanks{%
University of Leeds}}

\begin{document}
\maketitle

\begin{abstract}
The structure of a protein is crucial in determining its functionality, and is much more conserved than sequence during evolution. A key task in structural biology is to compare protein structures in order to determine evolutionary relationships, estimate the function of newly-discovered structures, and predict unknown structures. We propose a Bayesian method for protein structure alignment, with the prior on alignments  based on functions which penalise ``gaps'' in the aligned sequences. We show how a broad class of penalty functions fits into this framework, and how the resulting posterior distribution can be efficiently sampled. A commonly-used gap penalty function is shown to be a special case, and we propose a new penalty function which alleviates an undesirable feature of the commonly-used penalty. We illustrate our method on benchmark data sets, and find it competes well with popular tools from computational biology. Our method has the benefit of being able to potentially explore multiple competing alignments and quantify their merits probabilistically. The framework naturally allows for further information such as amino acid sequence to be included, and could be adapted to other situations such as flexible proteins or domain swaps. 
\end{abstract}

\small{Keywords:  Gap penalty prior, Markov chain Monte Carlo, Protein structure alignment, Structural bioinformatics, Unlabelled shape analysis.} 

\section{Introduction}\label{sec:gapintro}
\subsection{Background}\label{sec:background}

%\subsection{Motivation}

A protein is a chain of amino acids (of which there are $20$ types) that folds into a $3$-dimensional structure determined by the physical and chemical properties of the constituent amino acids. A crucial task in bioinformatics is to align a given pair of protein structures (such that, informally, they are ``as close as possible''), in order to quantify their similarity. This has many important applications, including inferring evolutionary relationships between proteins, predicting the functionality of newly-determined structures, and aiding protein structure prediction via template modelling or threading.

There have been many algorithms developed by computational biologists for this task, which have proved very successful and useful for particular aspects of the problem. These methods are optimised for particular sub-tasks, for example, suited to detecting more distant relationships or comparing particular aspects of structure, and are based on more heuristic scoring functions suited for the particular sub-task. These methods also typically assume the existence of a single, correct, alignment, and focus on returning this optimum. However, there may be alternative alignments of biological interest \citep{godzik96,shih04}, for example due to repeating substructures \citep{Mayr2007}. Additionally, there is inherent uncertainty due to errors in determining atomic coordinates experimentally and the natural vibration of proteins, which are not static. Inferences based on alignments should account for uncertainty in alignments as well as model parameters, for example, when using alignments to build phylogenies \citep{Sela2015}. Such approaches have been relatively well studied for protein sequence alignments \citep{Sela2015,Redlings2005}, but less well so for structural alignments \citep{Herman2019}. The aim of this paper is to place structural alignment in a fully probabilistic, Bayesian, framework, which deals with uncertainty in a robust and principled manner, and enables the possibility of alternative alignments to be explored and evaluated. We use \emph{sequence order} information to define a prior distribution on alignments, using functions which penalise ``gaps'' (in sequences) in an alignment; this extends the work of \citet{rodriguez14}, who use a particular gap penalty, to allow for more general penalty functions. We illustrate the idea by proposing a particular penalty function which encourages ``proportionality'' in alignments, but many other forms of penalty would fit into the framework. We also note that other sources of information, such as amino acid type, are readily accommodated in our framework; the focus of the paper is the form of penalty functions on which the prior distribution over alignments is based.

At the primary level, a protein is a sequence of letters from an alphabet $\mathcal{S}$ of $20$ letters representing the $20$ amino acids (sometimes referred to as residues). The most basic way of quantifying protein similarity is based on aligning the protein sequences. Consider a pair of proteins $S^x = \{s^x_j\}_{j=1}^{m}$ and $S^y = \{s^y_k\}_{k=1}^{n}$, consisting of $m$ and $n$ amino acids respectively, with elements $s^x_j,s^y_k \in \mathcal{S}$. The alignment task is to determine which amino acids correspond, or match, on each protein. \emph{Sequence} alignment methods use only the amino acid sequence to align the proteins. Given a mechanism for scoring matches between the different amino acids, plus a penalty for gaps, the ``best'' alignment can then be found by optimising this score over all possible alignments, which gives a measure of similarity between proteins. A widely-used method for this is BLAST \citep{altschul90}. See \citet{durbin98} for more details, and for probabilistic methods for protein sequence alignment based on hidden Markov models. \citet{zhu98} and \citet{liu99} discuss Bayesian sequence alignment. 

Figure \ref{fig:seqalignex} (a) shows an example of an alignment between two short protein sequences. In this example, the sequences are of the same length, and all amino acids are matched. Assuming, as we do throughout this paper, that the ordering of amino acids must be preserved in any alignment, then there is only one possible alignment. (This assumption is suitable when aligning \emph{homologous} structures, i.e. those evolved from a common ancestor.) In some positions, there is an alignment between identical amino acid types, and in other positions different amino acid types are aligned. In terms of an evolutionary model, this is viewed as a substitution, or mutation, and substitutions of amino acids with similar biological properties are more likely to occur. ``How likely'' can be measured via a similarity score for each pair of amino acids, where the score depends on the supposed evolutionary distance between the proteins. Measures such as PAM and BLOSUM matrices (see \citet{durbin98}) achieve this, essentially giving a log odds score for each pair relative to random mutations. An overall measure of alignment quality is then the sum of scores over all aligned pairs. A ``better'' alignment can be obtained by allowing some amino acids to not match, which can be achieved by introducing gaps in one or both sequences, as in Figures \ref{fig:seqalignex} (b) and (c). This allows more matches between the same, or biologically similar, amino acid types to be made. In evolutionary terms, gaps represent insertions or deletions (\emph{indels}) with respect to one of the sequences.

\begin{figure}[!h]
%\begin{minipage}{0.45\textwidth}
\begin{center}
\begin{tabular}{c|cccccccccc}
$S^x$&&G& K&S&T&L&L&K&K &L \\
$S^y$ &&G&K&G&T&I&C&K&A &L \\
\end{tabular}\\
\vspace{0.2cm}
(a) \\
\vspace{0.2cm}
\begin{tabular}{c|ccccccccccc}
$S^x$&&H&E&A&G&A&W&G&H&E&E \\
$S^y$&&P&-&-&-&A&W&H&E&A&E \\
\end{tabular}\\
\vspace{0.2cm}
(b)\\
\vspace{0.2cm}
\begin{tabular}{c|cccccccccccc}
$S^x$&&H&E&A&G&A&W&G&-&H&E&E \\
$S^y$&&-&P&-&-&A&W&-&A&H&E&E \\
\end{tabular}\\
\vspace{0.2cm}
(c)
\end{center}
%\end{minipage}\hfill
%\begin{minipage}{0.45\textwidth}
%\includegraphics[angle=270,scale=0.3]{1gky.ps}\\
%\includegraphics[angle=270,scale=0.3]{/maths/staff/pmzcf/LEEDSHOME/phd/scale/paper/revision/figs/2VLWsse2.ps}\\
%\end{minipage}
\caption{Three examples of a sequence alignment. In (a), there are no gaps. In (b) and (c) there are gaps (``-'') in one or both sequences, and of different lengths.} \label{fig:seqalignex}
\end{figure}
\begin{figure}[!h]
\begin{center}
\includegraphics[angle=270,scale=0.3]{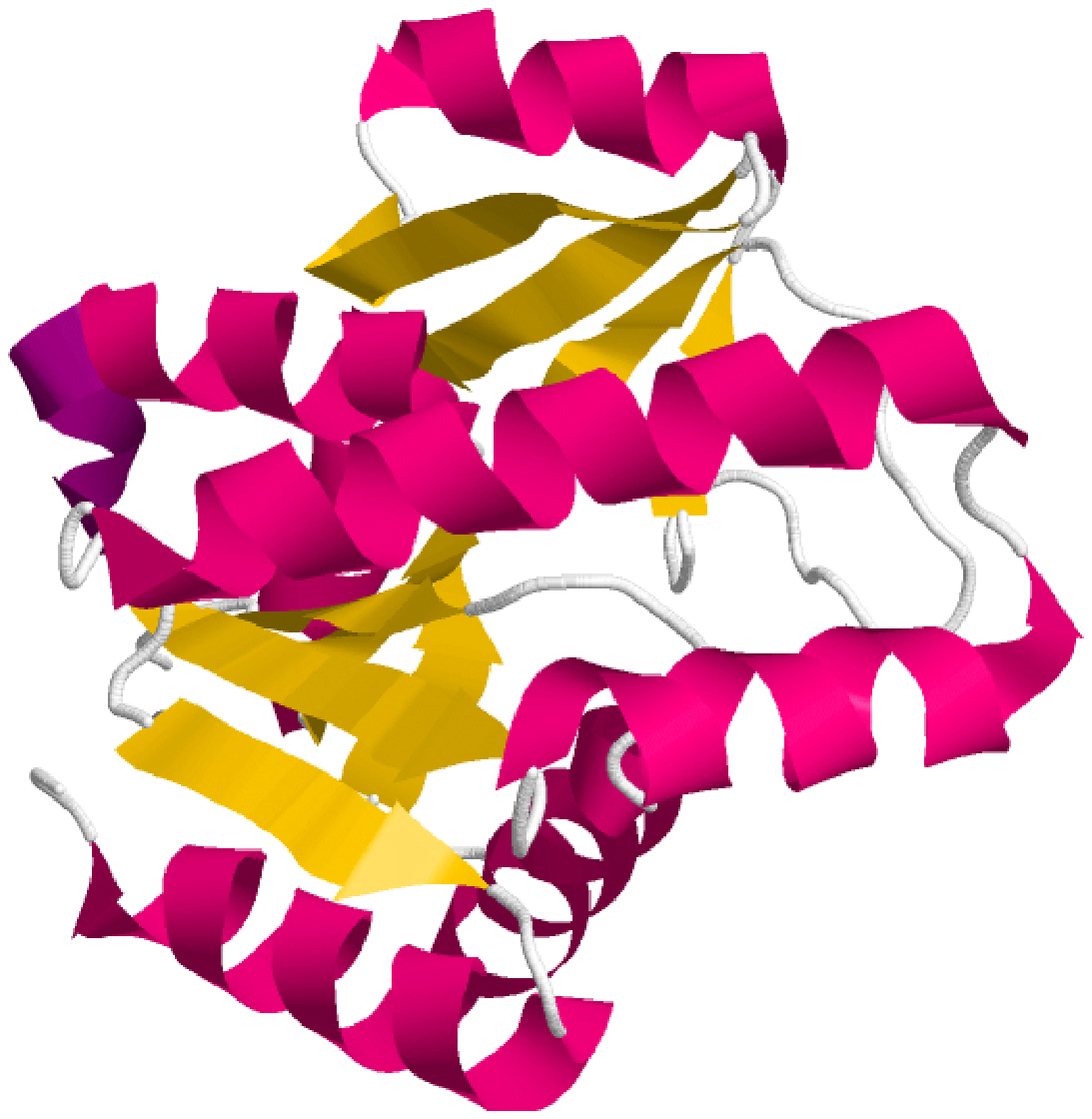}\\
\end{center}
\caption{An example protein structure. The secondary structure elements are the alpha helices and the beta strands (represented by arrows); the strands can be seen to lie in parallel to form structures called beta sheets.}\label{fig:protstruct}
\end{figure}

%For identifiability when displaying aligned sequences, we do not allow gaps in the $y$ sequence to follow immediately after gaps in the $x$ sequence
Subsections of the amino acid chain form structural units, called secondary structure elements (SSEs), the main two being alpha helices and beta strands (which can then form beta sheets). The sections of the protein chain between the SSEs are known as loops; the spatial arrangement of the secondary structure is called the tertiary structure, and is key to a protein's functionality. An example of a protein tertiary structure is given in Figure \ref{fig:protstruct}. Since the structure of a protein is more conserved than its sequence, protein structure alignment is more informative than sequence alignment. Over a period of evolution, the sequence of a protein may change through substitutions of amino acid residues from one type into another at a particular position, from the insertion of new amino acid residues, or from deletion of existing residues. However, the overall physical structure may remain essentially unchanged, at least in regions of the protein which are functionally important. Therefore, a better measure of how closely two proteins are related can be obtained by aligning their structures, and with the increasing number of protein structures becoming available and deposited in databases such as the Protein Data Bank (PDB, \citet{berman00}), reliable methods for protein structure alignment are crucial in protein bioinformatics --- see \citet{MardiaRSS13} for more background.

In this paper, we describe a fully Bayesian model for the alignment of proteins using structural information in the form of atomic coordinates of the amino acids, and hence the problem is one of statistical shape analysis (in particular, unlabelled shape analysis (e.g. \citet{GM,Dryden2016,rodriguez14}). We use \emph{sequence order} in the modelling, in the sense that the sequence order information forms the basis of our prior distribution over all possible alignments, through the use of gap penalty functions. (Note however that our model can also easily incorporate other sources of sequence information, such as amino acid type.)  Our prior distribution uses a penalty function of very general form, and hence the framework allows for a very rich and flexible class of prior distributions which can capture desirable features of an alignment. The framework allows the use of a penalty function commonly used in bioinformatics as a special case, on which \citet{rodriguez14} based their prior distribution on alignments. However, this penalty function has some undesirable properties, and its widespread use can arguably be attributed to its simplicity and ease of computational implementation rather than its biological realism. Here, we propose one possible penalty function which fits into our general framework, motivated by a desirable ``proportionality'' property (see Section \ref{sec:propprior}) which is very natural and plausible biologically. We note that there are many other possibilities which fit into this framework, and hence our model could be used in other contexts, with penalties chosen to capture particular desirable features related to the problem at hand.             

The key ingredients of our methodology are:
\begin{itemize}
\item A probability model for the atomic coordinates, conditional on a particular amino acid correspondence (alignment) and geometric transformation (Section \ref{sec:Lik}).
\item A prior distribution over all possible alignments which retain sequence order, based on a gap penalty function. We show how a general class of penalties fit into the framework (Section \ref{sec:gapprior}), and suggest a particular penalty function which addresses an intuitively undesirable feature of a popular penalty function (Section \ref{sec:propprior}). 
\item A posterior distribution over all alignments and transformation parameters (Section \ref{sec:gapprior}), such that all sources of uncertainty are handled in a principled manner, allowing alternative alignments to be explored and their merits quantified probabilistically. In Section \ref{sec:updateM} we show how alignments can be sampled efficiently from the posterior distribution, and that other priors of a very general form could be incorporated in the same way.    
\end{itemize}

\subsection{Protein similarity measures}\label{sec:ProtSim}

Once proteins have been aligned, it is often useful to quantify the similarity between them. These measures are also used to compare different alignments between the same proteins, and so optimising them can also be part of the alignment process. In this section, we discuss some of the issues in quantifying protein structure similarity, and some of the measures which have been proposed to do so. 

Consider two proteins $\bm{X}$ and $\bm{Y}$ with $m$ and $n$ residues respectively. Suppose $L$ residues have been aligned, where $\{\bx_i\}_{i=1}^L$ and $\{\by_i\}_{i=1}^L$ are the atomic coordinates of these points on $\bm{X}$ and $\bm{Y}$ respectively. A basic measure of similarity is the root mean squared deviation (RMSD), given by 
$$ \sqrt{\frac{1}{L}\displaystyle\sum_{i=1}^L ||\bx_i - \bA\by_i - \btau||^2}, $$ 
where $\bA$ is the rotation matrix and $\btau$ the translation vector giving the optimal superposition of $\bm{Y}$ on $\bm{X}$. A problem with interpreting RMSD as an absolute measure of similarity is that longer alignments (larger $L$) tend to have larger RMSD. Since long alignments and low RMSD are both desirable, there is a trade-off between choosing a longer alignment and smaller RMSD. This also makes comparing different alignments (e.g. those from different methods) difficult, unless one is Pareto optimal (longer alignment and smaller RMSD). In the Matt algorithm, \citet{menke08} and \citet{Daniels2012a} considered the balance of RMSD and $L$ as a proxy for homology, and found a linear combination of the two measures which achieved optimal performance on a particular benchmark data set, SABmark \citep{Sabmark04}, in terms of separating homologous proteins at superfamily level from decoys. 

Two measures which have been proposed as more general and robust measures of similarity are Global Template Score (GDT) \citep{zemla03} and Template Modelling (TM) score \citep{zhang04}. GDT is defined as $$100\frac{(n_1+n_2+n_4+n_8)}{4L},$$
where $n_a$ is the (cumulative) number of aligned residues which are less than $a ~ \mbox{\normalfont\AA}$ apart after superposition of the proteins. TMscore is defined as 
$$ \frac{1}{m}\displaystyle\sum_{i=1}^L \frac{1}{1+(\frac{d_i}{d_0})^2},$$
where $d_0 = 1.24 \sqrt[3]{m-15}-1.8$, $d_i$ is the distance between the $i^{\rm{th}}$ pair of residues after superposition, and $m$ is the total number of residues on the smaller protein.  The parameter $d_0$ provides a scale for normalising distances, and the TMscore was designed to improve upon measures of the same form, but which use constant values of $d_0$, such as $3.5$ or $5$. For measures which use a fixed value of $d_0$, \citet{zhang04} found a power law relationship between measurement scores and length (of the smaller protein) for randomly-chosen protein pairs with $< 30\%$ sequence similarity. The value of $d_0$ used in TMscore depends on the length of the smaller protein, and was found to correct the observed bias towards longer proteins in other measures.

\subsection{Review of existing methods}\label{sec:bioinfmethods}
%A summary of the key ideas is as follows:
%\begin{itemize}
%\item We show how the general framework of GM for matching point sets can be adapted to \emph{sequence order dependent} matching, thus providing a fully probabilistic model for protein structure alignment capable of incorporating various sources of information and uncertainty in a robust and coherent manner, and exploring/quantifying different alignments.
%\item Building on the prior suggested by rodriguez, we show how more general priors can be easily accommodated, and suggest a particular prior which alleviates one undesirable feature of the prior of rodriguez (which is derived from a very common gap penalty function used in bioinformatics).
%Section ....
%\item Demonstrate competitive performance with popular methods from computation biology on benchmark data sets, illustrating the utility of probabilistic methods based on a fully coherent probabilistic principles whilst retaining biological relevance. (Section results) 
%\end{itemize}
Here, we briefly review the main approaches for representing, and some of the algorithms developed for aligning, protein structures. The list is far from exhaustive, and we focus on covering the main approaches and some of the most well-known/popular methods. For a review of the main ideas and approaches to structure alignment, see \citet{hasegawa09} and \citet{Ma14}, who also give a comprehensive account of the available methods and discuss some open challenges. 

The most common representations of protein structures used in pairwise structure comparison methods are as sets of points, distance matrices, or secondary structure elements. The first approach represents the protein by a set of atomic coordinates, typically the $C_\alpha$ (alpha-carbon) atoms (which is also the approach we adopt in this paper). The second approach uses a matrix of intra-protein distances, such as the matrix of distances between $C_\alpha$ atoms. The final approach uses SSEs (alpha helices and beta strands). Some methods use a combination of these, or incorporate additional knowledge such as amino acid type and hydrogen bonding information. 

TMalign \citep{zhang05}, LGA  \citep{zemla03}, Matt \citep{menke08}, FATCAT \citep{Ye03}  and DeepAlign \citep{wang13} are examples of methods using the point set representation. TMalign and LGA consider rigid-body motions in order to establish an optimal alignment. Matt and FATCAT consider flexibility in the proteins by identifying fragments of structure, with ``bends'' allowed between the fragments. Formatt \citep{Daniels2012} builds on Matt by including amino acid sequence information to refine structural alignments. DeepAlign also utilises additional information from the amino acid sequence, as well as hydrogen bonds and local substructure substitution scores. DALI \citep{holm93} uses the distance matrix representation, and CE \citep{shindyalov98} uses a combination of atomic locations and distances. The VAST algorithm \citep{gibrat97} uses the SSEs, represented as vectors.  

Computational methods such as these seek to optimise some heuristically-justified scoring function, with the optimisation taking place over amino acid correspondences and/or some form of geometrical superposition.  In this paper, we introduce a fully probabilistic model which incorporates all sources of uncertainty in a principled manner, allowing alternative alignments to be explored and their merits quantified probabilistically.

The overall structure of the paper is as follows.  In Section \ref{sec:unlabelled} we formulate the problem as one of unlabelled shape analysis, and briefly discuss some alternative statistical methods. In Section \ref{sec:BayesStruct}, we describe the Bayesian model for structural alignment, and give details of our new prior distribution over alignments. In Section \ref{sec:Gapexamples} we apply our method to some challenging examples and benchmark data sets, and compare with some popular structural alignment methods, before concluding with a discussion.

\section{Mathematical formulation: unlabelled shape analysis}\label{sec:unlabelled}

In this section, we formulate the problem as one of (unlabelled) statistical shape analysis, and review some of the relevant literature in this area. 
 
Mathematically, a protein can be represented as a configuration  of $m$ points, $\{\bx_j\}_{j=1}^m$, $\bx_j \in \mathbb{R}^3$. For example, the points often represent the locations of the $C_\alpha$ (alpha-carbon) atoms of the amino acids. The problem is then to align this configuration with that of another protein $\{\by_k\}_{k=1}^n$. That is, we seek a rigid body transformation such that 
$$  \bA\by + \btau = \bx $$   
for any pair of points $\bx$ and $\by$ which are ``matched'' --- i.e. $\bx$ and $\by$ are equivalent points on their respective configurations. Here, $\bA$ is a $3 \times 3$ rotation matrix and $\btau \in \mathbb{R}^3$ is a translation vector. The correspondence between points on the two configurations is encoded in an $m \times n$ matrix $\bM$, with elements $M_{jk}$, where 
\begin{center}
\begin{math}
M_{jk}=\left\{\begin{array}{cc} 1, &  \textrm{if} \hsp \bx_j \hsp \textrm{and} \hsp \by_k \hsp \textrm{are matched,} \\0, &  \textrm{otherwise.}
\end{array}\right.
\end{math}
\end{center} 
Usually, $\bM$ is not known and it is the main object of interest about which to draw inference; this is known as unlabelled shape analysis, and the problem of protein structure alignment is an important example of this.

Unlabelled shape analysis has been the focus of much recent research interest in statistical shape analysis, motivated by important applications such as that of protein structure alignment. From the Bayesian viewpoint, there are essentially two approaches that have been developed for unlabelled shape analysis. One approach is to maximize over the transformation parameters $\bA$ and $\btau$ \citep{dryden07,schmidler07,rodriguez14} which   can be viewed as  using a Laplace approximation to integrate out $\bA$ and $\btau$ and using the marginal posterior distribution for inference about $\bM$  \citep{kenobi2012}. An alternative approach is to consider a fully Bayesian model, where the transformation parameters are included as unknown parameters in the model about which to draw inference \citep{GM}. In this manner, uncertainty in these parameters is accounted for and correctly propagated throughout the analysis \citep{wilkinson07}. Other approaches to the unlabelled shape alignment problem include the Softassign Procrustes method of \citet{rang97} and methods using the EM algorithm \citep{kent10,myronenko10}. A closely-related problem is that of looking for instances of a known shape in cluttered point clouds, for example searching for shapes in noisy images \citep{srivastava09,su13}.

An alternative description of a protein useful for statistical modelling is as a sequence of pairs of dihedral angles defined by the backbone of a protein, with each pair a point on a torus. For example, \citet{Boom_etal08} used hidden Markov models as generative models for protein structure.    \citet{lennox09} use Dirichlet process mixtures for density estimation of the distribution of dihedral angles, a problem also considered by \citet{maadooliat16} using nonparametric density estimation, focussing on modelling loop regions of a protein; see also \citet{Najibi17}. Motivated by direct modelling of the evolution of a protein, \citet{golden17} describe the shape of a protein as a sequence of dihedral angles on the torus; their  model captures dependencies between sequence and structure evolution through a diffusion process on the torus.

In this paper, we consider the alignment of protein structures within the fully Bayesian framework of \citet{GM}, but with an important change to the prior model for the matching matrix $\bM$. In the original setting of \citet{GM}, conditional on the total number of matched points $L$, then every possible $\bM$ consistent with $L$ matched points was considered equally likely. When aligning homologous proteins which are thought to have evolved from a common ancestor, it is important to preserve the sequence order of the points in the matching given by $\bM$. Therefore, we require a prior for $\bM$ which imposes this constraint. This has previously been considered by \citet{rodriguez14}, who worked out in detail the case corresponding to a prior based on a commonly-used penalty function in bioinformatics, which they used in their applications; they also suggested that more general priors, applicable in other situations, could be incorporated in this framework. We introduce a class of priors based on a more general penalty function, which alleviates the unappealing feature that, conditional on the numbers of matches and gaps, the indices of the points forming the matches are independent under the prior model. We show how this new prior can be incorporated into the fully Bayesian framework of \citet{GM}, and how an MCMC scheme can be implemented in light of the changes to the model due to this prior. (See also \citet{MardiaRSS13}, who considered alignment preserving sequence order, but with a uniform prior over all possible such alignments.) This methodology can give biologically-meaningful alignments on challenging problems, as well as accounting for uncertainty in the alignment and transformation parameters in a fully Bayesian manner. 

%Our suggested prior extends that of \citet{rodriguez14}, using a prior based on a commonly-used penalty function in bioinformatics.who used a prior based on a commonly-used penalty function in bioinformatics. 

The underlying formulation is very flexible. For instance, \citet{GM} considered rigid body transformations in their applications, but \citet{mardia13} demonstrated applications using full similarity transformations.  \citet{forbes14} also use this approach with similarity transformations in the context of fingerprint matching. \citet{green15} describes how the MAD-Bayes technique \citep{broderick13} can be used to obtain approximations to the MAP (maximum a-posteriori) estimator, useful when very fast approximate solutions might be needed in practical situations using very large data sets, a problem also considered by \citet{schmidler07}.

\section{Bayesian structure alignment}\label{sec:BayesStruct}
            
We now describe our Bayesian model for protein structure alignment. In Section \ref{sec:Lik} we form the likelihood, and in Section \ref{sec:gapprior} describe a general form of prior distribution on alignments based on gap penalty functions, showing how a commonly-used penalty is a special case. In Section \ref{sec:propprior}, a new prior for the matching matrix $\bM$ is proposed, which alleviates an undesirable property of the aforementioned common penalty function. In Section \ref{sec:updateM}, we detail our sampler for drawing samples from the posterior distribution of alignments, and in Section \ref{sec:Mest} describe how a point estimate alignment is obtained.  

\subsection{Likelihood}\label{sec:Lik}

We have two point configurations, $\bm{X} = \{\bx\}$ and $\bm{Y} = \{\by\}$, consisting of $m$ and $n$ points respectively. The points are labelled $\bx_j, \hsp j=1,\ldots,m$ and $\by_k, \hsp k=1,\ldots,n$, where $\bx_j ,\hsp \by_k \in \mathbb{R}^d$; in our case, protein structures are $3$-dimensional configurations and $d=3$. A rigid body transformation which transforms points on $\{\by\}$ into $x$-space is of the form $\bA\by + \btau$, where $\bA$ is a $d \times d$ rotation matrix and $\btau \in \mathbb{R}^d$ is a translation vector. As in \citet{GM}, we have 
\begin{eqnarray}\nonumber
 \bx_j & = & \bmu_{\xi_j} + \bepsilon_j \hsp \hsp j=1,\ldots,m, \\\nonumber
\bA\by_k + \btau & = & \bmu_{\eta_k} + \bepsilon_k \hsp \hsp k=1,\ldots,n, 
\end{eqnarray}
where $\{\bmu\}$ is an unobserved hidden configuration, from which the observed points are derived.  The $\bepsilon$ terms represent error in the observed points, which are regarded as noisy observations of the true locations on $\{\bmu\}$. Here, we use a spherical Gaussian model for the errors, so that $\bepsilon \sim N_d(0,\sigma^2 \bm{I})$, where $\bm{I}$ is the $d \times d$ identity matrix; the parameter $\sigma^2$ therefore represents the error variance. The $\xi$ and $\eta$ terms give the mapping between points on $\{\bmu\}$ and points on $\{\bx\}$ and $\{\by\}$ respectively. In particular, when $\xi_j = \eta_k$ then the corresponding $\bx$ and $\by$ points are both realisations of the same hidden location, and are regarded as matched points. The matching between the configurations is captured by the matching matrix $\bM$. We impose the constraint that a given point on one configuration can match at most one point on the other configuration, so that each row or column of $\bM$ has at most one non-zero entry. Then, $\displaystyle\sum_{j,k} M_{jk} =L$, where $L$ is the total number of matched pairs of points.

The points on $\{\bmu\}$ are assumed to form a homogeneous Poisson process over a region of volume $v$, and these hidden points can be integrated out. Then, assuming $v$ is large relative to the support of the density of the error terms, the (approximate) respective likelihood contributions of the unmatched $\bx$, unmatched $\by$ and matched points are 
$$ v^{-(m-L)}, \hsp (|\bA|/v)^{n-L}, \hsp (|\bA|/v)^{L} \displaystyle\prod_{j,k:M_{jk}=1} \frac{\phi\{(\bx_j -\bA\by_k -\btau)/(\sigma \sqrt 2)\}}{ (\sigma \sqrt 2)^d} ,$$
where $\phi(\cdot)$ is the $d$-dimensional standard normal density. Hence the likelihood of the observed data given $\bM$ (and the other parameters) is

\begin{equation}\label{eq:gaplik}
p(\bx,\by|\bM,\bA,\btau,\sigma) = v^{-(m+n-L)} |\bA|^n \displaystyle\prod_{j,k:M_{jk}=1} \frac{\phi\{(\bx_j -\bA\by_k -\btau)/(\sigma \sqrt 2)\}}{ (\sigma \sqrt 2)^d}.
\end{equation}

\subsection{Gap prior}\label{sec:gapprior}

Recall that our main objective is to align two configurations when the points on each configuration have a meaningful ordering which must be preserved in any resulting alignment, which may include gaps in the corresponding sequence alignment in one or both of the sequences. We summarise an alignment with the matching matrix $\bM$, for which we use a prior which imposes the sequence order constraint. As a starting point, we use the prior 
\begin{equation}\label{eq:gapprior}
p(\bM;g,h) = Z(g,h)\exp\{-u(\bM;g,h)\}, 
\end{equation}
as in \citet{rodriguez14}, where $u(\bM;g,h)$ is a penalty function which penalises gaps in the alignment, and $Z(g,h)$ is a normalising constant. The parameters $g$ and $h$ are known as gap opening and extension penalties respectively. The penalty function is  
 \begin{equation}\label{eq:gappen}
%u(\bM;g,h) = gS(\bM) + h\sum_{i=1}^{S(\bM)} (l_i -1)
u(\bM;g,h) = gs(\bM) + hl(\bM)
\end{equation}
where $s(\bM)$ is the number of instances where a new gap in the alignment is opened, $l(\bM)=\sum_{i=1}^{S(\bM)} (l_i -1)$, and $l_i$ is the length of the $i$th gap. This corresponds to a gap penalty function widely used in sequence alignment \citep{durbin98}. The prior (\ref{eq:gapprior}) using penalty function (\ref{eq:gappen}) was used by \citet{liu99} in a Bayesian sequence alignment setting. \citet{wu98} also used the penalty (\ref{eq:gappen}), in what we believe to be the first statistical work on protein sequence and structure alignment, which was set in a regression framework. 

To illustrate what is meant by a new gap and length of a gap, consider again the sequence alignment in Figure \ref{fig:seqalignex} (b). In the first sequence, the second residue is not matched to a residue on the second sequence; instead it is aligned to a ``-'', indicating that a gap has been opened. That is, a gap opening is said to have been created where a residue in one sequence is unmatched, but the previous residue in the same sequence was aligned to a residue in the other sequence. The length of the gap is then the number of unmatched residues (in the same sequence) until another matched pair; therefore, the gap in Figure \ref{fig:seqalignex} (b) is of length $3$.

In Figure \ref{fig:seqalignex} (c), the first sequence has one gap, of length $1$, and the second sequence has three gaps, of lengths $1$, $2$ and $1$. Note that the two sequences are considered independently when counting the number and length of the gaps, so that a gap in one sequence followed immediately by a gap in the other sequence would be counted as two different gap openings.

Before introducing a generalisation of the prior distribution (\ref{eq:gapprior}), we first illustrate how this prior fits into our framework. Recall that configurations $\bm{X}$ and $\bm{Y}$ consist of $m$ and $n$ points respectively, and suppose that there are $L$ matched points between the two. Further, suppose the indices of the matched points on $\bm{X}$ are  $j_0 < j_1 < j_2 < \cdots < j_L < j_{L+1} $ and the indices of the matched points on $\bm{Y}$ are $k_0 < k_1 < k_2 < \cdots < k_L < k_{L+1}$.  Hence, if $j_{i+1} - j_i \ge 2$, there is a gap in the $\bm{X}$ sequence of length $j_{i+1} - j_i -1 $, and similarly for $\bm{Y}$ involving the $k$ indices. We set $j_0 = k_0 =0 $ and $j_{L+1} = m+1$, $k_{L+1}=n+1$, which are artificial matching indices, fixed throughout, introduced to account for the start and end points of the sequences.

Hence, the total penalty given by (\ref{eq:gappen}) is 
$$ u(\bM;g,h) = \displaystyle\sum_{i=0}^{L} f(j_{i+1} - j_{i}) + \displaystyle\sum_{i=0}^{L} f(k_{i+1} - k_{i}), $$
where
\begin{equation}\nonumber f(r) = \left\{\begin{array}{cc}
0 & \hspace{0.4cm} r = 1 \\
g & \hspace{0.4cm} r=2 \\ 
g + (r-2)h & \hspace{0.4cm} r > 2. \\
\end{array}\right.
\end{equation}  
Thus, the total penalty can be easily computed as a sum of simple contributions involving consecutive pairs of the matched point indices. In the same spirit, we can generalise the prior distribution (\ref{eq:gapprior}) by incorporating other penalty functions $u(\bM;\bm{\phi})$ which are expressible as a sum of penalty contributions involving small subsets of the matching indices. Here, $\bm{\phi}$ is a vector of parameters, and in the special case (\ref{eq:gappen}), we have $ \bm{\phi} = (g,h)$.  This has positive implications for the implementation, as follows. The MCMC sampling methods we employ (described in Section \ref{sec:updateM}) involve computing differences of the form $u(\bM^{\prime};\bm{\phi}) - u(\bM;\bm{\phi})$, where $\bM^{\prime}$ is a proposed modification of $\bM$. The computation will be efficient if the change from $\bM$ to $\bM^{\prime}$ only affects a small number of the terms which comprise $u(\bM;\bm{\phi})$, and each of these terms are simple to compute. We describe a novel penalty function in Section \ref{sec:propprior} that adheres to this principle, which corresponds to a prior which can control the degree of ``proportionality'' in the indices of the matched points. 

%$u(\bM;\bm{\phi})$ is expressible as a sum of contributions which each involve only a small number of the matched point indices, and only a small number of these contributions are affected by the change from $\bM$ to $\bM^{\prime}$.  

%$$ u(\bM;g,h) = \displaystyle\sum_{i=0}^{L} f(j_{i+1} - j_{i}) + \displaystyle\sum_{i=0}^{L} f(k_{i+1} - k_{i}), $$
%   The penalty in (\ref{eq:gappen}) is a special case of this formulation, with
%\begin{equation}\nonumber f(r) = \left\{\begin{array}{cc}
%0 & \hspace{0.4cm} r = 1 \\
%g & \hspace{0.4cm} r=2 \\ 
%g + (r-2)h & \hspace{0.4cm} r > 2. \\
%\end{array}\right.
%\end{equation}
%Note also that using this formulation, other penalty functions could be incorporated in the same way. In particular, any function which can be decomposed into a sum of penalty contributions involving subsets of the matching indices could be used. We describe a novel penalty function in Section \ref{sec:propprior}, which corresponds to a prior which can control the degree of ``proportionality'' in the indices of the matched points.  
%The function penalises a gap opening and the subsequent length of the gapIt is common to penalise gap openings more than extensions, i.e. set $g > h$.
Since we are using a different form of prior distribution on $\bM$ to that considered by \citet{GM}, there is a minor change to the joint model (Equation (6) in that paper). As described above, the priors we consider are of the general form
\begin{equation}\label{eq:genprior}
p(\bM;\bm{\phi}) \propto \exp\{-u(\bM;\bm{\phi})\}. 
\end{equation}
Multiplying (\ref{eq:gaplik}) and (\ref{eq:genprior}), we obtain 
$$ p(\bM,\bx,\by|\bA,\btau,\sigma) \propto |\bA|^n v^{L}  \exp \{-u(\bM;\bm{\phi})\} \displaystyle\prod_{j,k:M_{jk}=1} \frac{\phi\{(\bx_j -\bA\by_k -\btau)/(\sigma \sqrt 2)\}}{ (\sigma \sqrt 2)^d} $$
and the joint model is  
\begin{equation}\label{eq:gapjoint}
\begin{split}
 p(\bM,\bA,\btau,\sigma,\bx,\by)  & \propto  p(\bA)p(\btau)p(\sigma)|\bA|^n v^{L}  \exp \{-u(\bM;\bm{\phi})\} \\ 
    & \quad \times  \displaystyle\prod_{j,k:M_{jk}=1} \frac{\phi\{(\bx_j -\bA\by_k -\btau)/(\sigma \sqrt 2)\}}{ (\sigma \sqrt 2)^d}.\\
\end{split}
\end{equation}
In particular, the term $v^L$ remains, unlike in the model of \citet{GM}, where this term cancelled with a corresponding term from the prior for $\bM$. We discuss specification of $v$ in our applications in Section \ref{sec:settings}.  The prior distributions on $\bA$, $\btau$ and $\sigma$ are $p(\bA),p(\btau)$ and $p(\sigma)$ respectively. The rotation matrix $\bA$ has a matrix-Fisher prior distribution, where $p(\bA) \propto \exp\left\{ \mbox{tr} (\bm{F}_0^T\bA)\right\}$ and the parameter $\bm{F}_0$ is a $d \times d$ matrix.  $\bA$ is parametrised  by Eulerian angles, $\theta_{12},\theta_{13},\theta_{23},$ say, in the case $d=3$. In our examples we use a uniform prior on $\bA$, which is the special case where $\bm{F}_0$ is the $d \times d$ matrix of zeroes. $\bA$ then has a uniform prior with respect to the invariant measure on $SO(3)$, the Haar measure, where $SO(3)$ is the special orthogonal group of all $d \times d$ rotation matrices. For the translation vector $\btau$, we have $\btau \sim N_d(\bmu_{\tau},\sigma_{\tau}^2 \bm{I}_d)$, where $\bmu_{\tau}$ is a mean vector and $\sigma_{\tau}^2 \bm{I}_d$ a covariance matrix, with $\bm{I}_d$ the $d \times d$ identity matrix. For the noise parameter $\sigma$, we have $\sigma^{-2} \sim \Gamma(\alpha,\beta) $, so $p(\sigma^{-2}) \propto \sigma^{-2(\alpha-1)}\exp\left(-\frac{\beta}{\sigma^{2}}\right) $. 

%Note that the volume term $v$ is now no longer absorbed into the normalising constant, unlike in the model of \citet{GM}, where this term cancelled with a corresponding term from the prior for $\bM$. We discuss sensitivity to user-specified values of $v$ in Section \ref{sec:vsens}.

\subsection{A proportionality prior}\label{sec:propprior}

We now describe our new penalty function, which controls ``proportionality'' in the alignment and contains the penalty in (\ref{eq:gappen}) as a special case. 

Consider the pair of triples $(j_1,j_2,j_3)$ and $(k_1,k_2,k_3)$, from which we obtain the pair $(j_2-j_1)$,  $(j_3-j_2)$ from the $\bm{X}$ sequence and the pair  $(k_2-k_1)$ , $(k_3-k_2)$ from the $\bm{Y}$ sequence. Given  $j_1,j_3,k_1,k_3$, we would prefer $j_2$ and $k_2$ such that the ratio 
$$ \frac{(j_2-j_1)/(j_3-j_2)}{(k_2-k_1)/(k_3-k_2)} $$
is close to one. 

In general, given $L$ matches, we have $L$ triples of matching indices in the $\bm{X}$ sequence, given by  \\ $(j_0,j_1,j_2),(j_1,j_2,j_3),\ldots,(j_{L-1},j_L,j_{L+1})$. Similarly, in the $\bm{Y}$ sequence we have the $L$ triples $(k_0,k_1,k_2),(k_1,k_2,k_3),\ldots,(k_{L-1},k_L,k_{L+1})$. For the $i$th pair of triples, consider the log ratio 
$$ q_i = \log\left\{ \frac{(j_{i}-j_{i-1})/(j_{i+1}-j_{i})}{(k_{i}-k_{i-1})/(k_{i+1}-k_{i})} \right\} .$$    
%Then a Gaussian penalty on the proportionality term is given by $-\log[\phi(q_i/\sigma_q)/\sigma_q] $, where $\phi(\cdot)$ is the probability density function of the standard normal distribution and $\sigma_q$ is a specified parameter controlling tolerance to log ratios different to zero. In other words, the penalty is given by 
%$$ \gamma(q_i;\sigma_q) = \frac{1}{2}\log(2\pi) + \log \sigma_q + \frac{q_i^2}{2 \sigma_q^2} .$$
Then a Gaussian-type penalty on the lack of proportionality is given by
$$\gamma(q_i;\nu) = \frac{\nu q_i^2}{2}.$$
Combining this with the penalty function (\ref{eq:gappen}) (which uses only $s(\bM)$ and $l(\bM)$), the total penalty function is 
$$u(\bM;g,h,\nu) = gs(\bM) + hl(\bM) + \displaystyle\sum_{i=1}^L \gamma(q_i;\nu) .$$ 
Letting $\nu=0$, we obtain the original penalty (\ref{eq:gappen}).

For example, consider the case with $m=8$, $n=17$ and $L=3$. Two possible alignments ($\bM_1$ and $\bM_2$ respectively say) are 
$$
\begin{array}{cccccc}
\bM_1:     & j_0 & j_1 & j_2 & j_3 & j_4 \\
 & 0 & 2 & 5 & 7 & 9 \\
 & 0 & 4 & 10 & 14 & 18 \\
 & k_0 & k_1 & k_2 & k_3 & k_4 
\end{array}
 $$
and
$$
\begin{array}{cccccc}
\bM_2: & j_0 & j_1 & j_2 & j_3 & j_4 \\
&0 & 2 & 5 & 7 & 9 \\
&0 & 2 & 12 & 16 & 18 \\
 & k_0 & k_1 & k_2 & k_3 & k_4 
\end{array}
 $$
In both cases, $s(\bM) = 8$ and $l(\bM) = 11$ and hence the original gap penalty is the same, so $p(\bM_1;g,h)/p(\bM_2;g,h) = 1$ under the original gap penalty prior. 
%$\frac{p(\bM_1;g,h)}{p(\bM_2;g,h)} = 1$

Consider now the prior with the penalty for lack of proportionality included. For $\bM_1$, we have $q_1 = q_2 = q_3 = 0$ (all ratios are equal to $1$). This gives a total penalty of $ 8g + 11h $, the same as the original gap penalty.
For $\bM_2$ we have :
$$
\begin{array}{c}
q_1 = \log\left\{ \frac{(j_1-j_0)/(j_2-j_1)}{(k_1-k_0)/(k_2-k_1)}\right\}  = \log\left(\frac{2/3}{2/10}\right) = 1.204.
\end{array}
$$
With $\nu = 1$, this gives a penalty of $\gamma(1.204;1) = 0.5\times1.204^2 = 0.725 $ for the first pair of triples. 
Similarly, $q_2 = \log(0.60) = -0.511$, giving a penalty of  $0.131$, and $q_3 = \log(0.5) = -0.693$, resulting in a penalty of $0.240$. The total penalty is 
$$   8g + 11h + 0.725 + 0.131 + 0.240 = 8g + 11h + 1.096 .$$
Hence, under the new prior,  $p(\bM_1;g,h,\nu)/p(\bM_2;g,h,\nu) = \exp(1.096) = 2.99 $.

Note that larger values of $\nu$ penalise a lack of proportionality more. For instance, in the example above with $\nu=4$ we have 
$$  \frac{p(\bM_1;g,h,\nu)}{p(\bM_2;g,h,\nu)} = \exp(4.38) = 80 $$
under the new prior, so $\bM_1$ (which preserves proportionality perfectly) is  strongly preferred over $\bM_2$.

\subsection{Sampling $\bM$}\label{sec:updateM}

Updates for the parameters $\bA$, $\btau$ and $\sigma$ are as in \citet{GM}. We now describe the mechanism for generating posterior samples of $\bM$, using Metropolis-Hastings updates. Suppose our current alignment is $\bM$, and we have a proposal value $\mpr$ drawn from a proposal density $q(\mpr;\bM)$. Then the acceptance probability is 
$$ \alpha = \mbox{min} \left\{1,\frac{p (\mpr,\bA,\btau,\sigma,\bx,\by)q(\bM;\mpr)}{p (\bM,\bA,\btau,\sigma,\bx,\by)q(\mpr;\bM)} \right\},$$ 
where $p(\cdot)$ is the joint model (\ref{eq:gapjoint}). 

Similar to \citet{GM}, we consider three types of update for $\bM$, namely adding a matched pair, deleting a matched pair, or switching a matched pair, but the form of the updates is different due to the new prior on $\bM$.  We illustrate the idea by considering adding a matched pair, and the other two cases are similar; full details of our sampler are given in supplementary information. Suppose there are currently $L$ matches with indices $j_1 < j_2 < \cdots < j_L $ and $k_1 < k_2 < \cdots < k_L $. Suppose further that we propose to add a match ($j^*$,$k^*$), where $j_i < j^* < j_{i+1}$  and $k_i < k^* < k_{i+1} $, $i = 0,\ldots,L$, and we also have $j_0 = k_0 = 0 $ and $j_{L+1} = m +1 , k_{L+1} = n+1$. 
Then 
$$\frac{p (\mpr,\bA,\btau,\sigma,\bx,\by)}{p (\bM,\bA,\btau,\sigma,\bx,\by)} = \exp\{u(\bM;\bm{\phi}) - u(\mpr;\bm{\phi})\}\times \frac{v \phi\{(\bx_{j^*} -\bA\by_{k^*} -\btau)/(\sigma \sqrt 2)\}}{ (\sigma \sqrt 2)^d},$$
where $u(\bM;\bm{\phi}) - u(\mpr;\bm{\phi})$ is the reduction in the gap penalty achieved by adding the match ($j^*$,$k^*$). As described in Section \ref{sec:gapprior}, the penalty functions we consider are of a form which facilitates efficient computation of this reduction; since only a small number of terms involving matched indices either side of ($j^*$,$k^*$) are affected, it is not necessary to recalculate the whole penalty each time a change to $\bM$ is proposed. 

Note that under this sampling method, we make only small perturbations to the alignment at each iteration, by either removing a match, adding a match, or switching a match, so that the total number of matches can change by at most $1$. Our sampler is quite simple compared to that used by \citet{rodriguez14}, who propose global changes to $M$ using dynamic programming recursions analogous to those used in sequence alignment algorithms \citep{zhu98,liu99}, which may improve performance. Instead, we improve performance of our sampler, which makes local changes to $M$, using parallel tempering \citep{geyer91}.

%In contrast to that used by \citet{rodriguez14}, our sampler is quite simple, and we improve performance using the generic method 

%This method of sampling $\bM$ from the posterior distribution is quite different to the method used by \citet{rodriguez14}. These authors sample a whole new matching matrix $\bM$ at each iteration of the MCMC sampler, using dynamic programming recursions analogous to those used in sequence alignment algorithms. Therefore, generating proposal moves for $\bM$ is more computationally intensive than in our case, but the sampler may converge to equilibrium faster than our method which uses local moves, so that fewer iterations of the sampler are needed overall. 
\subsection{Point estimation}\label{sec:Mest}

%The RMSD is calculated for a particular alignment, or value of $\bM$. For example, we could calculate the RMSD for a particular estimate of $\bM$, $\hat{\bM}$ say. 
As described in \citet{GM}, the principles of Bayesian decision theory can be used to obtain a posterior point estimate for $\bM$ from our MCMC output, by defining a loss function which incorporates costs for falsely declaring matches and missing true matches.  It is necessary only to specify a value for a parameter $K$, where $K = c_{01}/(c_{01}+c_{10})$; the term $c_{01}$ denotes the cost incurred for falsely declaring a match, and $c_{10}$ denotes the cost of falsely missing a true match.  The point estimate is obtained by minimising the expected loss with respect to the marginal posterior matching probabilities, which can be regarded as a linear assignment problem. Note that larger values of $K$ give fewer matches, since falsely declaring a match incurs a relatively higher cost than missing a true match. As a default, we take $K=0.5$, so that both types of error are considered equally costly. To solve the linear assignment problem, we use the method of \citet{jonker87}.

\section{Applications}\label{sec:Gapexamples}

%We now illustrate the new methodology with two examples, also analyzed by \citet{rodriguez14}, which have been studied previously in the literature. These are challenging examples, where in each case the proteins have a low sequence identity (percentage of aligned pairs which are the same amino acid residue type), but are structural homologues. In our first example, we analyse one of these pairs, with PDB identification codes 1ACX and 1COB. In our second example, we analyse the pair of proteins with PDB identification codes 1GKY and 2AK3. We then give results of a larger-scale comparison, using $16$ protein pairs considered to be challenging proteins for structural alignment methods. These pairs were also considered by \citet{rodriguez14}, who also compared with the CE algorithm; we show that our method produces results which improve on these methods in many cases.   %  also analyzed by \citet{rodriguez10}.

We now illustrate our methodology with applications to real protein data. We first analyse the pair of proteins with PDB identification codes 1GKY and 2AK3, also analysed by \citet{rodriguez14}, which have been studied previously in the structural bioinformatics literature.  We then investigate the performance of our method on a set of $16$ protein pairs considered to be challenging for structural alignment methods \citep{jung2000,ortiz02}, which \citet{rodriguez14} used to compare their method against the CE algorithm of \citet{shindyalov98}.  We then present results on the MALIDUP and RIPC benchmark data sets, and compare with some of the available methods from computational biology. In all cases, we find that our results are competitive with these other methods.

%  and show that our results are very competitive alongside CE and those reported by  \citet{rodriguez14}.
%all $16$ of these pairs, and show that our results are very competitive alongside CE and those reported by  \citet{rodriguez14}. 

%also analyzed by \citet{rodriguez14}, which have been studied previously in the structural bioinformatics literature. These are challenging examples, where in each case the proteins have a low sequence identity (percentage of aligned pairs which are the same amino acid residue type), but are structural homologues (i.e. the proteins have evolved from a common ancestor). In our first example, we analyse the pair of proteins with PDB identification codes 1GKY and 2AK3. Our second example is the pair 1ACX-1COB, one of $16$ protein pairs considered to be challenging for structural alignment methods, which \citet{rodriguez14} used to compare their method against the CE algorithm. We then summarize the results using our method on all $16$ of these pairs, and show that our results are very competitive alongside CE and those reported by  \citet{rodriguez14}. 

\subsection{Parameter settings}\label{sec:settings}

It is necessary to specify a value of the parameter $v$, which represents the volume of the region in which the configurations of points are realised. We specify $v$ as follows. Let $\bar{\Omega}_x = \prod_{i=1}^{3} \{\max_j(x_{ji}) - \min_j(x_{ji})\} $ be the volume of the region containing the $\bm{X}$ configuration. Similarly, let $\bar{\Omega}_y = \prod_{i=1}^{3} \{\max_k(y_{ki}) - \min_k(y_{ki})\}.$ Then define $\bar{\Omega} = \max\{\bar{\Omega}_x,\bar{\Omega}_y \}.$ As a default, we take $v = 1.2 \bar{\Omega}$, which is the value used for the reported results. We found our results to be robust to increases in this parameter.

%It is necessary to specify a value of the volume parameter $v$. We find that $v$ can have a marked effect on the posterior number of matches, with higher values of $v$ favouring more matches. From experience, we find that a value of $v=5000$ provides a good starting point in practice, generally giving a reasonable number of matches and sensible solutions (as evidenced, for example, by comparing the resulting alignments with those from a deterministic alignment program, such as LGA \citep{zemla03}). The value of $v$ can then be adjusted to promote more or fewer matches as desired; in Section \ref{sec:vsens}, we further investigate the effect of changing $v$ for both the examples in this section. However, it should be remembered that $v$ is not merely an abstract parameter, but is a parameter in the model which represents the  volume of the space in which the (hidden) points lie. Therefore, if real information is known about this, then it could be used to obtain a representative value of $v$ --- in our context, this could be the volume of the space which contains the larger of the two observed configurations. In practice, however, it could also be used as a tuning parameter by the user to experiment with a greater/lesser prior propensity for matching.     

The parameter settings we used for the user-defined parameters which remain fixed throughout this section are summarized in Table \ref{tab:paramsettings}. We use the values $g=4$ and $h=0.1$ for the gap opening and extension penalty parameters, reflecting that the opening of a gap should be penalised more than extending a gap, to discourage alignments with lots of short gaps which are not plausible biologically \citep{altschul88}. The values we use are equal to the expected values of $g$ and $h$ from the prior distributions used by \citet{rodriguez14}, who suggest that a gap opening penalty of the order of $40$ times as large as the gap extension penalty is reasonable, following \citet{gerstein98}.  For the parameter $\nu$, we compare the results obtained using the values $0.25$ and $4.0$ in order to assess the effect of our new prior on the resulting alignments. We chose these values after numerical experimentation using simulations from the prior distribution on $M$; we found $\nu=4.0$ encourages proportionality very strongly, to the point where this excessively dominates the gap opening and extension penalties, whereas a value of $\nu=0.25$ gave a more even balance between proportionality and gaps. More details, including plots of simulations from the prior, are given in the supplementary material.      
%\begin{table}
%\begin{center}
%\fbox{%
%\noindent\makebox[\textwidth]{
%\fbox{
%\begin{tabular}{*{5}{c}}
%$g$ & $h$ & $v$ & $\beta$ & $\sigma_{\tau}$ \\\hline
%$4$ & $0.1$ & $5000$ & $8$ & $500$  \\
%\end{tabular}
%}
%\caption{\label{tab:paramsettings} Parameter settings for  user-defined parameters which remain fixed.}
%\end{center}
%}
%\end{table}
\begin{table}
\caption{\label{tab:paramsettings}Parameter settings for user-defined parameters which remain fixed.}
\centering
%\noindent\makebox[\textwidth]{
%\fbox{
\begin{tabular}{|cccc|}\hline
$g$ & $h$ &  $\beta$ & $\sigma_{\tau}$ \\\hline
$4$ & $0.1$ &  $8$ & $500$  \\\hline
\end{tabular}
%}
%}
\end{table}

 %For both our examples, we also consider the case where $g$ and $h$ are treated as additional unknown parameters in our model, and apply the methods of Section \ref{sec:gapunkowns}. 
%\begin{table}
%\caption{\label{tab:paramsettings} Parameter settings for \\ user-defined parameters which remain fixed.}
%\centering
%\fbox{%
%\begin{tabular}{*{5}{c}}
%$g$ & $h$ & $v$ & $\beta$ & $\sigma_{\tau}$ \\\hline
%$4$ & $0.1$ & $5000$ & $8$ & $500$  \\
%\end{tabular}}
%\end{table}

For the remaining parameters, we use the following settings. The prior mean for the translation, $\bmu_{\tau}$, is taken to be the difference between the centroids of the two configurations. Prior information on $\btau$ is weak, so we set $\sigma_{\tau}=500$ to give a diffuse prior to reflect this. The prior for the rotation matrix $\bA$ is uniform. We set $\alpha = 1$, giving an exponential prior for $\sigma^{-2}$ with mean $1/\beta$. We keep $\beta =8 $ fixed throughout, corresponding to a mean precision broadly similar to the resolutions typical of PDB data (around $1$--$3$ \mbox{\normalfont\AA}) --- posterior inferences are robust to moderate changes of this value for $\beta$. The initial matching matrix $\bM$ was taken to be the zero matrix, corresponding to no matched points. All results relating to a specific alignment were obtained using the point estimate of $M$ with $K=0.5$, as discussed in Section \ref{sec:Mest}, which means false negative and false positive matches are equally undesirable.

With unlabelled shape analysis in general, the posterior distribution is known to be inherently multimodal, with the potential for MCMC samplers to become trapped in subsidiary modes \citep{dryden07,rodriguez14} corresponding to poor alignments. There may also be more than one genuinely-interesting mode, corresponding to different alignments of biological interest, and a strength of the Bayesian approach is the potential ability to explore the full posterior distribution and quantify the relative merits of each. To help ensure good convergence and mixing properties of the sampler, we used the parallel tempering method \citep{geyer91}, with $N=6$ chains at temperatures $T_1 < T_2 < \cdots < T_6$, where $T_1=1$ is the chain corresponding to the target posterior distribution and we used $T_6=32$. For the remaining temperatures, the following scheme was used: $T_i = (1/T_{i+1} + \Delta)^{-1}, i=2,\ldots,5$, where $\Delta = (1-1/T_6)/(N-1)$. This scheme was found via experimentation to perform well. Multiple chains were then run from different starting values for the parameters $\bA$, $\btau$ and $\sigma$, and posterior trace plots of the various parameters, as well as the log-posterior, were inspected visually. 

\subsection{Example}\label{sec:1GKY2AK3}

We first discuss alignment of the pair 1GKY (chain A, $186$ points) and 2AK3 (chain A, $226$ points). These are both kinases (enzymes which catalyze phosphorylation reactions); 1GKY in yeast and 2AK3 in cows.
The proteins have a low sequence identity (percentage of aligned pairs which are the same amino acid residue type) \citep{zhu98}, but are structural homologues (i.e. the proteins have evolved from a common ancestor) and are VAST structural neighbours \citep{zhu98,gibrat97}; hence, a structural alignment can detect this relationship, despite the low sequence similarity. We compare alignments obtained using two different values of $\nu$, namely $\nu=0.25$ (prior mean number of matches approximately $158$) and $\nu=4.0$ (prior mean number of matches approximately $148$).

\begin{figure}[!ht]
\begin{center}
\includegraphics[scale=0.57,angle=270,trim={0cm 0.3cm 2cm 0},clip]{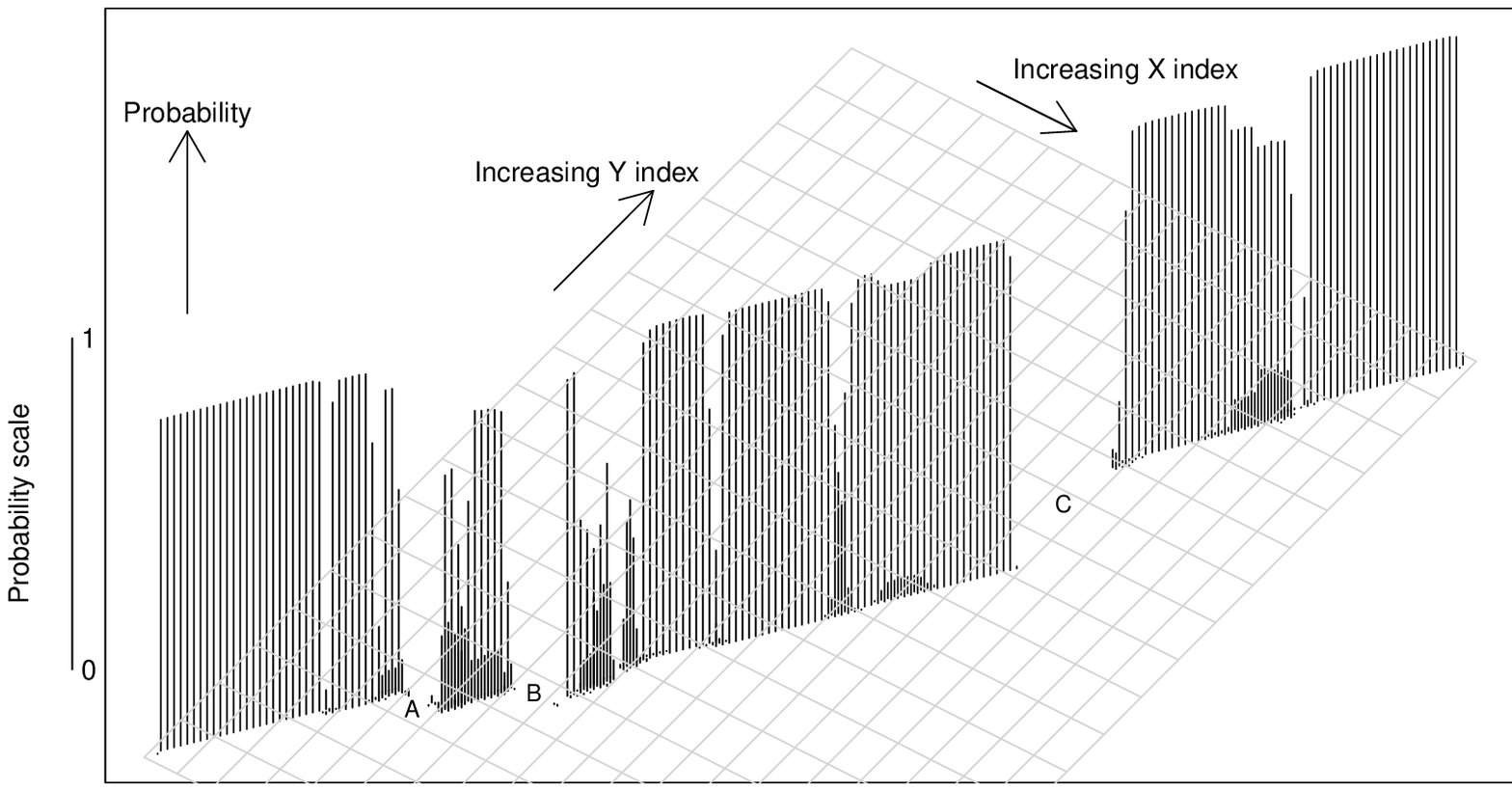}\\
(a)
\includegraphics[scale=0.57,angle=270,trim={0cm 0.3cm 2cm 0},clip]{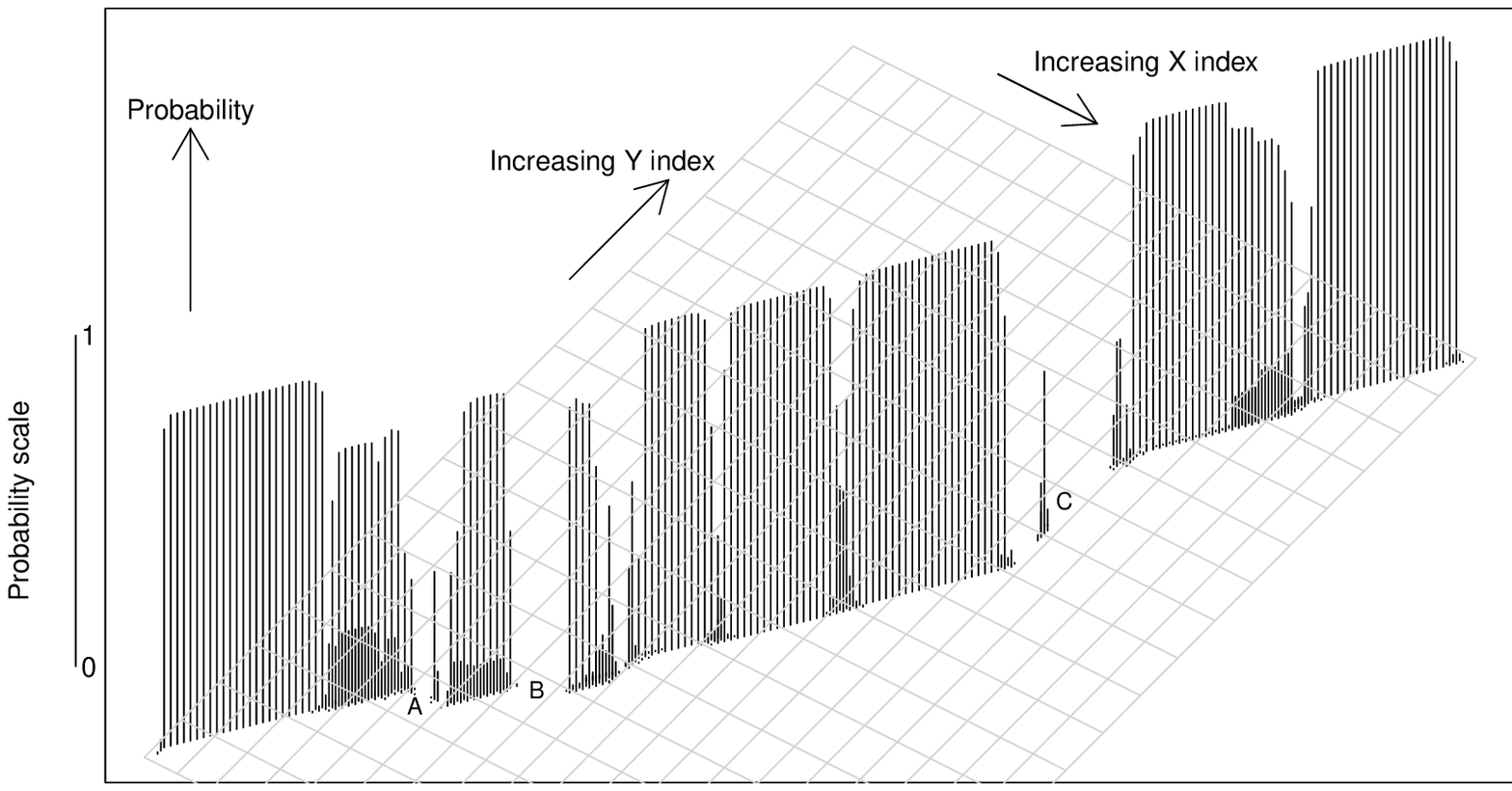}\\
(b)
\end{center}
\caption{Posterior matching probabilities between pairs of points on proteins 1GKY ($X$) and 2AK3 ($Y$), with  $\nu=0.25$ (a) and $\nu=4.0$ (b). The points are plotted in transformed coordinates to highlight the diagonal region of interest, where the horizontal coordinate is of the form $ax + by$ and the vertical coordinate is of the form $cy - dx$, where $x$ and $y$ are $X$/$Y$ indices and $c$ and $d$ are constants. The grid lines represent holding one of the original $X$/$Y$ coordinates fixed, with the direction of increasing $Y$/$X$ index indicated. The length of the vertical segment indicates the probability of the match between the corresponding pair of indices.}\label{fig:1GKY2AK3_comb20}
\end{figure}

Since the configurations contain $m$ and $n$ points, the array of pairwise posterior matching probabilities is of dimension $m \times n$. However, this array will be rather sparse, with the non-negligible probabilities concentrated around the diagonal due to the sequence order constraint. In order to display an alignment, we plot the posterior matching probabilities of pairs for which the probability exceeds $0.001$, and the axes are linear combinations of the indices chosen to clearly display the diagonal region of interest. Figure \ref{fig:1GKY2AK3_comb20} shows such displays for the two values of $\nu$ used. Each vertical segment corresponds to a matched pair, with the corresponding matching probability given by the length of the segment, the scale of which is indicated in the margin. The axes indicate the directions of increasing $\bm{X}$ (1GKY) and $\bm{Y}$ (2AK3) indices. For example, the regions marked A and B in Figure \ref{fig:1GKY2AK3_comb20} (a) indicate longer sections where points in 1GKY are not aligned to any points in 2AK3, and the region marked C indicates a longer section of  unaligned 2AK3 points. Figure \ref{fig:1GKY2AK3_comb20} (a) is a display of the matching probabilities for the case $\nu=0.25$. We clearly see sections of low uncertainty in the alignment, corresponding to conserved regions of structure which can be aligned very well, as well as regions where there is more uncertainty.  The point estimate $\hat{\bM}$ (using $K=0.5$) consists of $152$ matched pairs of points and a corresponding RMSD of $3.0$. 

Figure \ref{fig:1GKY2AK3_comb20} (b) shows the corresponding plot with $\nu=4.0$. Comparing with the previous alignment ($\nu=0.25$), the alignments tend to agree where there was low uncertainty, with any differences being in more uncertain regions, such as those directly preceding and following the regions marked A and B. Additionally, there is a small section of aligned points introduced in the region marked C. In this case, the point estimate $\hat{\bM}$ gives $153$ matched pairs of points and a corresponding RMSD of $3.2$. A value of $\nu=4.0$ penalises a lack of proportionality quite strongly --- by the analogy with a Gaussian distribution used to construct the penalty in Section \ref{sec:propprior}, $\nu$ is a precision parameter for the log ratio $q$, and $\nu=4.0$ corresponds to a standard deviation of $0.5$. Likewise, $\nu=0.25$ corresponds to a standard deviation of $2.0$. As a default, we use $\nu=0.25$, as used to obtain the following results presented in Sections \ref{sec:allPairs}, \ref{sec:Malidup} and \ref{sec:RIPC}.

\subsection{Comparison $1$: \citet{ortiz02} data set}\label{sec:allPairs}

We now present results on a set of $16$ pairs of proteins described by \citet{jung2000} and further analysed in \citet{ortiz02}. This set, which consists of protein pairs considered challenging for structural alignment methods, is used by \citet{rodriguez14} in their study. We shall refer to our method as SEQ-ALIBI (SEQuence-informed Alignment by Bayesian Inference), following ALIBI (\citet{MardiaRSS13}; see also \citet{GM}).  We compare the results of SEQ-ALIBI (SA)  with the method of \citet{rodriguez14}, using the results reported for their parameter $\lambda=8.6$ (RS), and those from the CE algorithm of \citet{shindyalov98}, as reported by \citet{rodriguez14}. The number of matches and corresponding RMSDs for each of the $16$ protein pairs are given in Table \ref{tab:RSproteins}. We use the previously-suggested default values of $\nu=0.25$ and $K=0.5$, and values of all other parameters are set as described in Section \ref{sec:settings}. A comparison of the results obtained using various other combinations of $\nu$ and $K$ can be found in the supplementary material. 

%We now show the results of a larger-scale comparison, using all $16$ pairs of proteins of \citet{ortiz02} and also used by \citet{rodriguez14} in their study. 
%We compare the results of SEQ-ALIBI, using $\sigma_q=2.0$ and $K=0.5$, with CE and the method of Rodriguez and Schmidler (2014) with their parameter $\lambda=8.6$ (RS). The number of matches and corresponding RMSDs for each of the $16$ protein pairs are given in Table \ref{tab:RSproteins}.
\clearpage

\begin{table}
\caption{\label{tab:RSproteins}RMSD and number of matches for the 16 protein pairs for CE,  RS  and SEQ-ALIBI.}
\centering
%\noindent\makebox[\textwidth]{
%\fbox{
\begin{tabular}{|c|c|c|c|c|c|c|c|}\hline
%\cline{1-6} & \multicolumn{6}{c}{K=0.5} \\ % & \cline{3-4} & \multicolumn{4}{c}{K=0.3} & \cline{5-6} & \multicolumn{2}{c}{K=0.1} \\
\multicolumn{1}{|c}{} & \multicolumn{1}{c|}{} & \multicolumn{2}{|c|}{CE} &  \multicolumn{2}{|c|}{RS} & \multicolumn{2}{|c|}{SEQ-ALIBI} \\\hline
Protein Pair & PDB IDs & RMSD & L & RMSD & L & RMSD & L \\\hline
1 & 1ABA-1DSB & 4.5 & 56 & 3.7 & 57 & 3.9 & 72 \\
2 & 1ABA-1TRS & 2.7 & 70 & 3.4 & 72 & 2.6 & 71 \\
3 & 1ACX-1COB & 4.0 & 92 & 3.8 & 86 & 2.7 & 84 \\
4 & 1ACX-1RBE & 7.3 & 56 & 2.8 & 31 & 4.3 & 52 \\
5 & 1MJC-5TSS & 2.7 & 61 & 3.0 & 60 & 2.3 & 60 \\
6 & 1PGB-5TSS & 2.9 & 48 & 3.3 & 55 & 2.7 & 55 \\
7 & 1PLC-1ACX & 3.3 & 80 & 4.0 & 84 & 2.8 & 73 \\
8 & 1PTS-1MUP & 4.1 & 80 & 3.1 & 83 & 2.8 & 85 \\
9 & 1TNF-1BMV & 4.1 & 115 & 4.2 & 109 & 3.7 & 112 \\
10 & 1UBQ-1FRD & 4.4 & 64 & 2.9 & 62 & 2.5 & 64 \\
11 & 1UBQ-4FXC & 4.0 & 64 & 2.9 & 61 & 2.6 & 64 \\
12 & 2GB1-1UBQ & 3.1 & 48 & 3.4 & 51 & 2.1 & 44 \\
13 & 2GB1-4FXC & 3.6 & 48 & 3.9 & 53 & 3.0 & 53 \\
14 & 2RSL-3CHY & 4.1 & 80 & 3.8 & 76 & 3.9 & 83 \\
15 & 2TMV-256B & 3.5 & 84 & 2.9 & 79 & 3.0 & 81 \\
16 & 3CHY-1RCF & 3.9 & 116 & 4.5 & 122 & 4.2 & 122 \\\hline
\end{tabular}
%}
%}
\end{table}

Table \ref{tab:SArelativeCE} summarizes the relative performance of CE and SEQ-ALIBI in terms of the trade-off between RMSD and number of matches. As discussed in Section \ref{sec:ProtSim}, typically one is improved at the expense of the other, making comparisons of different alignments difficult. However, longer alignments with lower RMSD (Pareto optimality) are clearly desirable. For $8$ of the $16$ pairs, SEQ-ALIBI finds Pareto optimal alignments, and on no occasion is the reverse true. 

\begin{table}
\caption{\label{tab:SArelativeCE}Comparison of number of matches ($L$) and RMSD between SEQ-ALIBI (SA) and CE for each of the $16$ protein pairs.}
\centering
%\noindent\makebox[\textwidth]{
%\fbox{
\begin{tabular}{|c|c|}\hline
 & Protein pair \\\hline
$L^{\mbox{SA}} \ge L^{\mbox{CE}} $ and $\mbox{RMSD}^{\mbox{SA}} < \mbox{RMSD} ^{\mbox{CE}}$ &  1  2  6  8 10 11 13 14  \\    
$L^{\mbox{SA}} \ge L^{\mbox{CE}} $ and $\mbox{RMSD}^{\mbox{SA}} > \mbox{RMSD} ^{\mbox{CE}}$ &  16  \\    
$L^{\mbox{SA}} < L^{\mbox{CE}} $ and $\mbox{RMSD}^{\mbox{SA}} > \mbox{RMSD} ^{\mbox{CE}}$ &  -  \\
$L^{\mbox{SA}} < L^{\mbox{CE}} $ and $\mbox{RMSD}^{\mbox{SA}} < \mbox{RMSD} ^{\mbox{CE}}$ &   3  4  5  7  9 12 15  \\\hline
\end{tabular}
%}
%}
\end{table}

\begin{figure}[!h]
\begin{center}
\includegraphics[scale=0.5,angle=270]{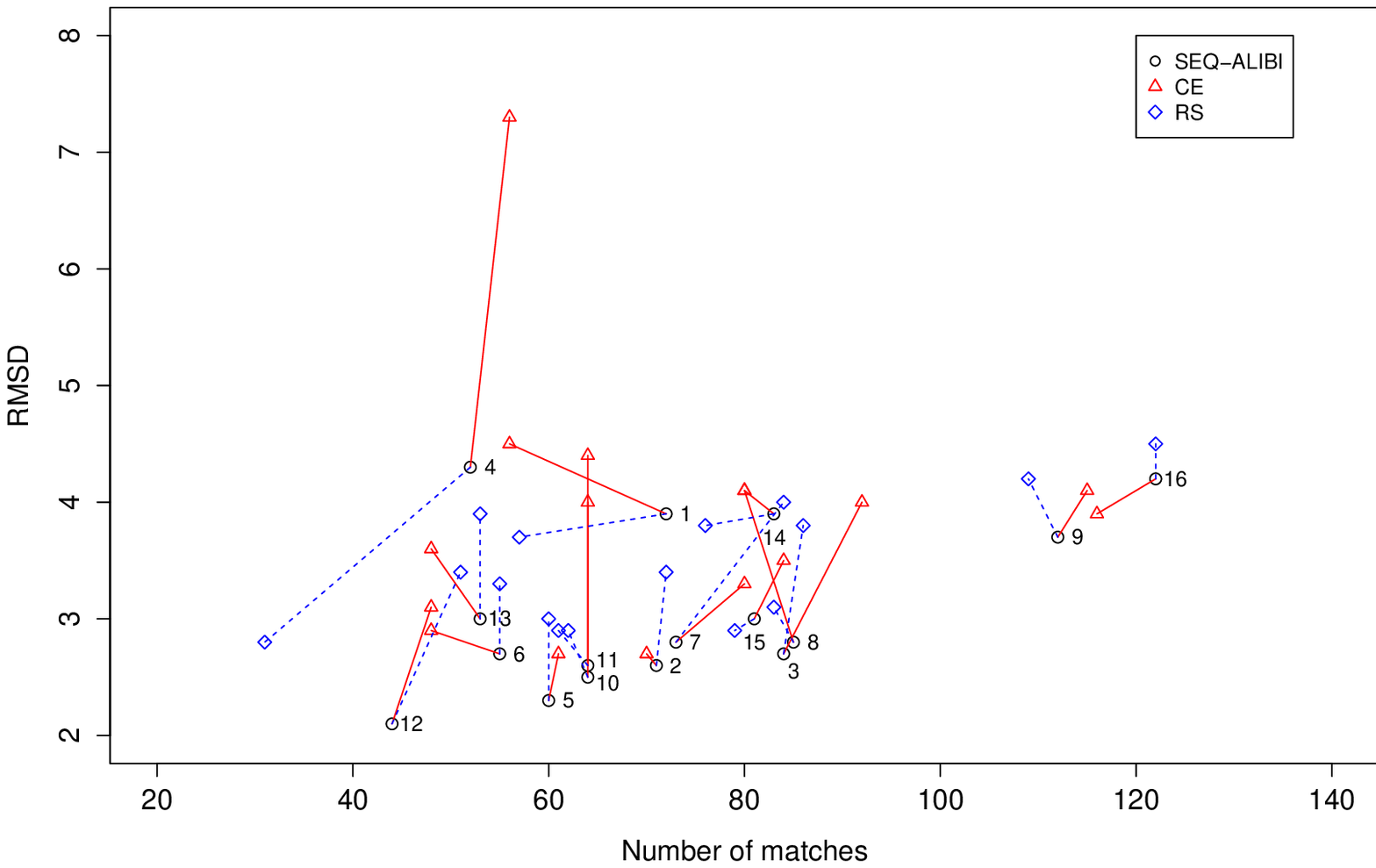}
\end{center}
\caption{RMSD against number of matches for each of the $16$ protein pairs using SA, CE and RS. The pairs are numbered as in Table \ref{tab:RSproteins}. For each pair, the line segments join the point for SA with the points for CE (solid line) and RS (dashed line).}\label{fig:SArelativeRS}
\end{figure}

Results from a similar comparison of SEQ-ALIBI with RS are given in Table \ref{tab:SArelativeRS}. Again, there are $8$ cases where SEQ-ALIBI finds Pareto optimal alignments, and none where the reverse is true. The results from all $3$ methods for all $16$ protein pairs are plotted in Figure \ref{fig:SArelativeRS}. For each pair, we have plotted RMSD against number of matches for each method, and drawn a line from the point for SEQ-ALIBI to the points for RS and CE. Near-vertical lines indicate an increase in RMSD for similar number of matches, and near-horizontal lines a reduction in matches for similar RMSD. Lines in a ``north-west'' direction signify the best performance of SEQ-ALIBI, since they indicate fewer matches and higher RMSD of the other method. Lines on the ``south-west--north-east'' axis correspond to pairs where neither method is clearly better, since more matches are being added at the cost of higher RMSD. Overall, we see that SEQ-ALIBI is at least as good as the other two methods, but does a lot better for some pairs.

\begin{table}
\caption{\label{tab:SArelativeRS}Comparison of number of matches ($L$) and RMSD between SEQ-ALIBI (SA) and RS for each of the $16$ protein pairs.}
\centering
%\noindent\makebox[\textwidth]{
%\fbox{
\begin{tabular}{|c|c|}\hline
 & Protein pair \\\hline
$L^{\mbox{SA}} \ge L^{\mbox{RS}} $ and $\mbox{RMSD}^{\mbox{SA}} < \mbox{RMSD} ^{\mbox{RS}}$ &   5  6  8  9 10 11 13 16  \\    
$L^{\mbox{SA}} \ge L^{\mbox{RS}} $ and $\mbox{RMSD}^{\mbox{SA}} > \mbox{RMSD} ^{\mbox{RS}}$ &  1 4 14 15  \\    
$L^{\mbox{SA}} < L^{\mbox{RS}} $ and $\mbox{RMSD}^{\mbox{SA}} > \mbox{RMSD} ^{\mbox{RS}}$ &  -  \\
$L^{\mbox{SA}} < L^{\mbox{RS}} $ and $\mbox{RMSD}^{\mbox{SA}} < \mbox{RMSD} ^{\mbox{RS}}$ &  2 3  7  12   \\\hline
\end{tabular}
%}
%}
\end{table} 

\subsection{Comparison $2$: MALIDUP data set}\label{sec:Malidup}

The MALIDUP (Manual ALIgnments of DUPlicated domains) \citep{cheng08} data set consists of $241$ protein pairs which can be used for a larger-scale comparison. Each pair has a corresponding reference alignment curated manually by experts. 

We compare results from our method, SEQ-ALIBI (SA), with the manual alignments and those from DALI \citep{holm93}, TMalign (TM) \citep{zhang05}, DeepAlign (DA) \citep{wang13} and Matt \citep{menke08}. 

\begin{figure}[!h]
\begin{minipage}{0.45\textwidth}
\begin{center}
\includegraphics[angle=270,scale=0.40]{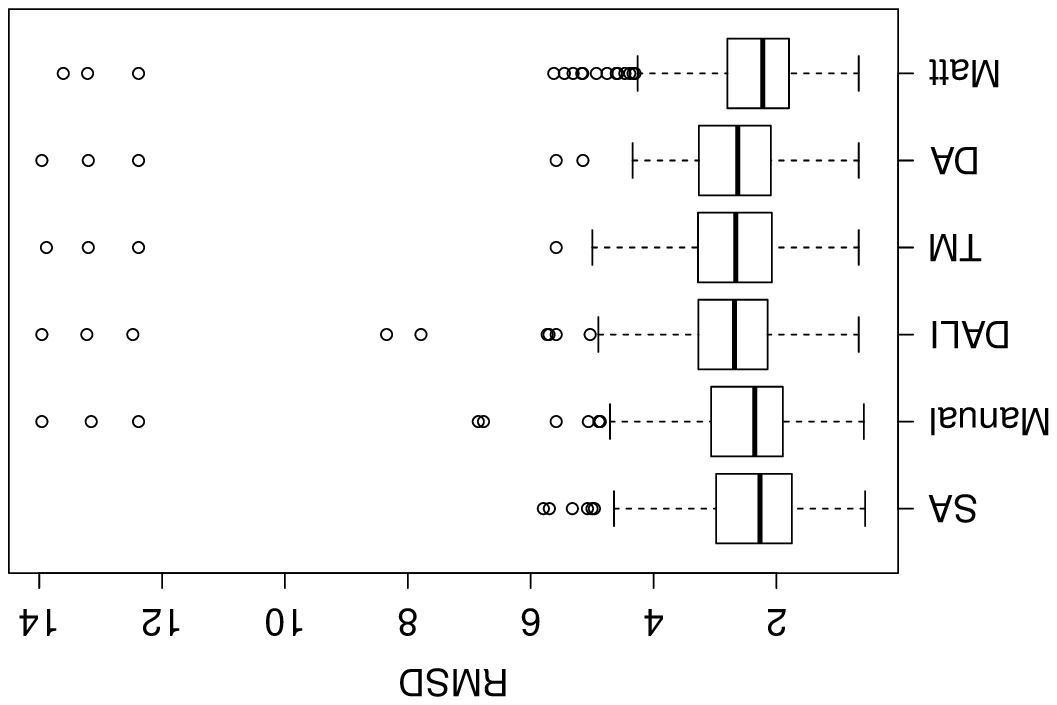}\\
\end{center}
\end{minipage}
\begin{minipage}{0.45\textwidth}
\begin{center}
\includegraphics[angle=270,scale=0.40]{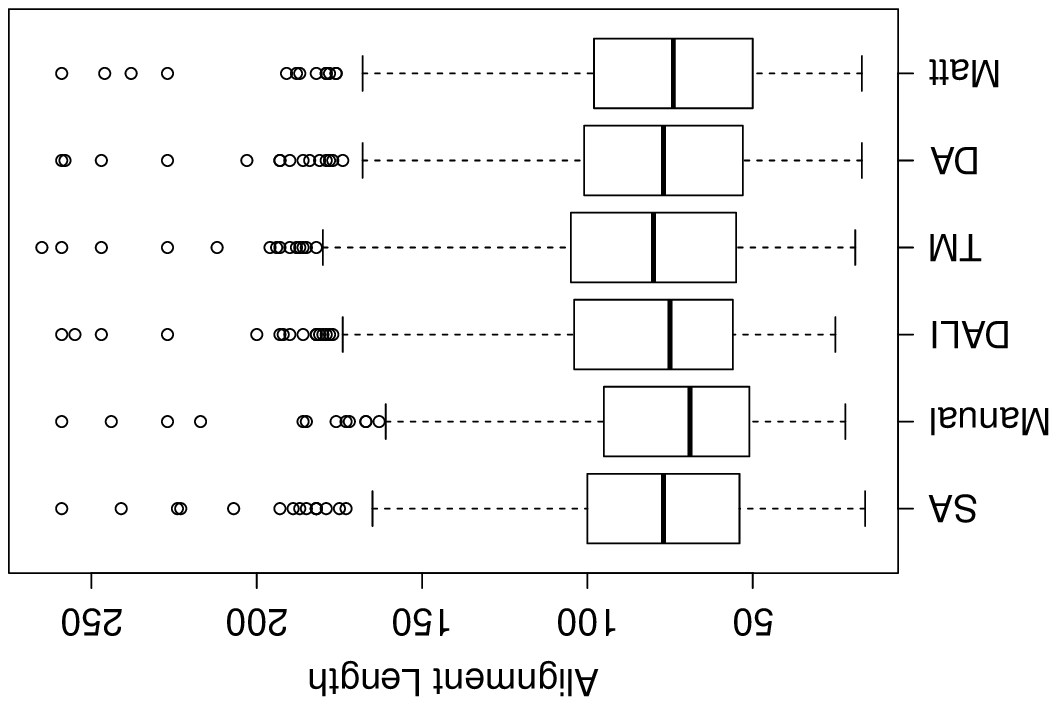}\\
\end{center}
\end{minipage}
\caption{RMSDs (left) and alignment lengths (right) for SA, manual, DALI, TM, DA and Matt.}\label{fig:MaliDupRMSD}
\end{figure}

We see (Figure \ref{fig:MaliDupRMSD}) that overall RMSD values are lower for SA than the other methods, except Matt, which is similar, but has shorter alignments generally (median 77 for SA and 74 for Matt). Overall, the distributions of alignment lengths are quite similar. As discussed in Section \ref{sec:ProtSim}, it is difficult to use these quantities as absolute measures of performance, but the distributions give a sense of the general picture. Note that for the $3$ outlying RMSD values present in the other methods, SA matches fewer points than the other methods, and these extra points increase the RMSD dramatically.

\begin{figure}[!h]
\begin{minipage}{0.3\textwidth}
\begin{center}
\includegraphics[angle=270,scale=0.4]{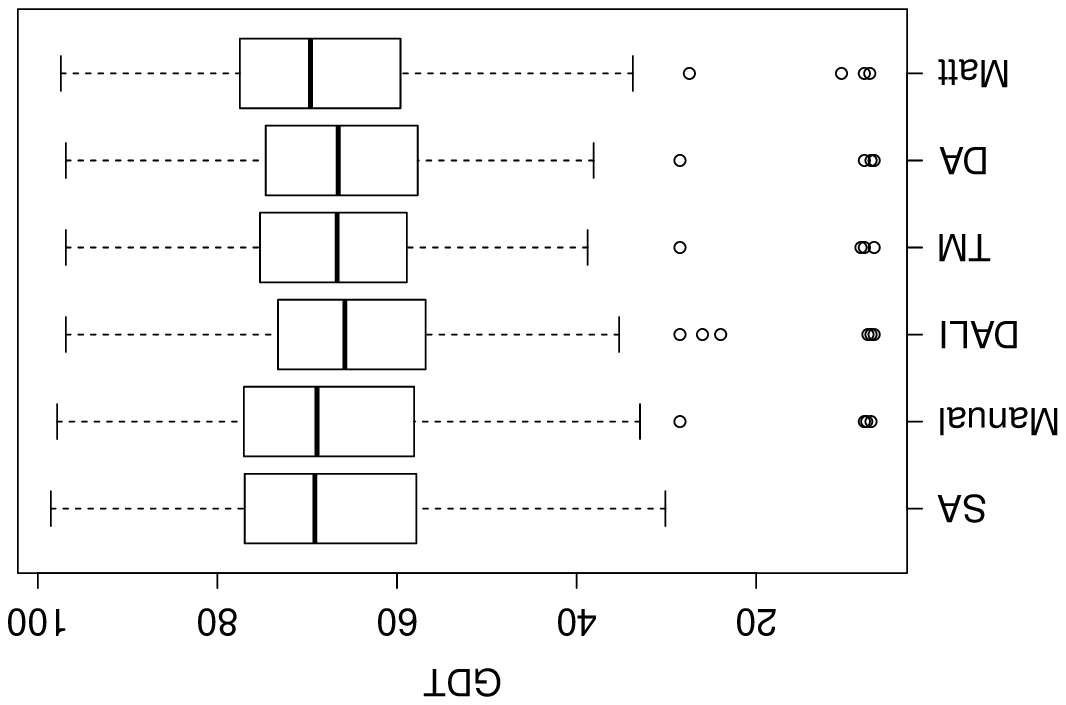}\\
\end{center}
\end{minipage}
\begin{minipage}{0.3\textwidth}
\begin{center}
\includegraphics[angle=270,scale=0.4]{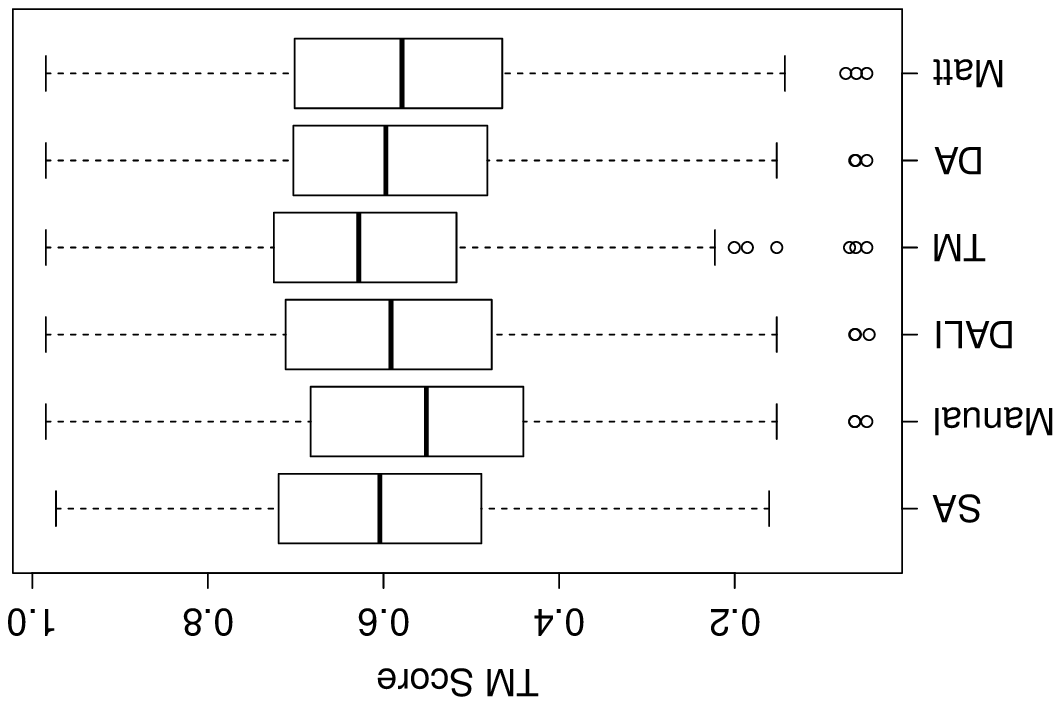}\\
\end{center}
\end{minipage}
\begin{minipage}{0.3\textwidth}
\begin{center}
\includegraphics[angle=270,scale=0.4]{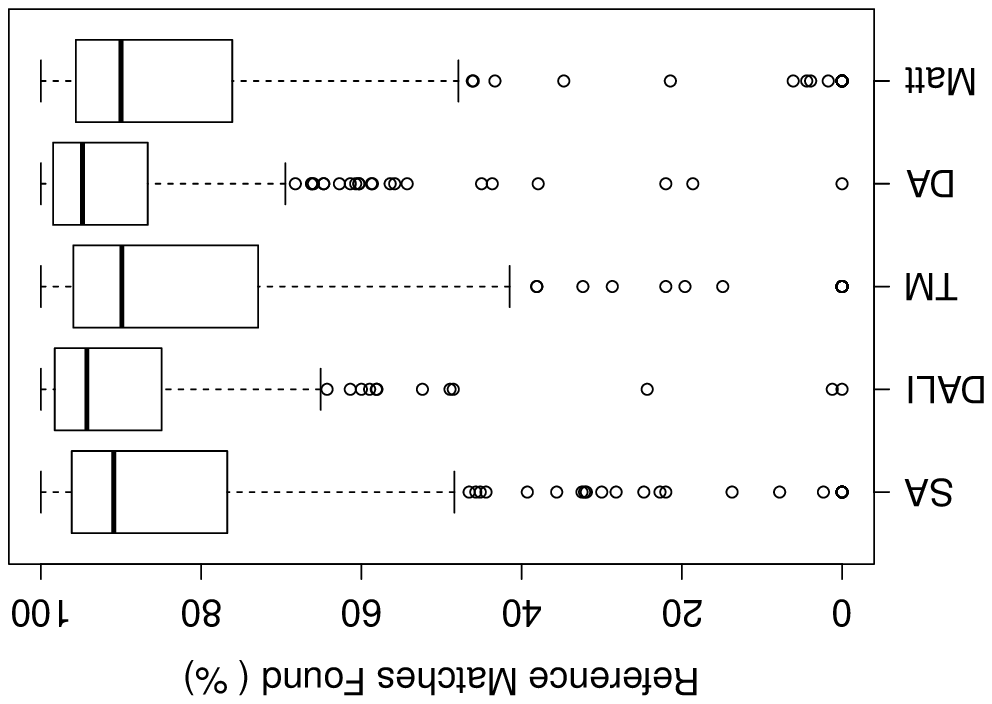}\\
\end{center}
\end{minipage}
\caption{GDT scores (left), TM scores (centre) and percentage of reference alignment matches found (right), for SA, manual, DALI, TM, DA and Matt on the MALIDUP data set.}\label{fig:MaliDupGDTTM}
\end{figure}

The distributions of GDT and TM scores can be seen in Figure \ref{fig:MaliDupGDTTM}. For GDT scores, we see that the distribution for SA is similar to that of the manual alignments and Matt, and higher than DALI, TM and DA. For TMscore, SA gives higher scores overall than the other methods, apart from TM, which is not surprising since TM optimises TMscore. 

%\begin{figure}[!h]
%\begin{center}
%\includegraphics[angle=270,scale=0.45]{Malidup_RefAlignBoxplot.ps}\\
%\end{center}
%\caption{Percentage of reference alignment matches found by SA, DALI, TM, DA and MATT.}\label{fig:MaliDupRef}
%\end{figure}

For agreement with the manual reference alignments, SA performs similarly, perhaps slightly better, than TM and Matt (Figure \ref{fig:MaliDupGDTTM}). DALI and DeepAlign perform best overall. We observe that, for the pairs where SA has lower agreement with the reference alignment, the SA alignments tend to be longer and have lower RMSD than the manual alignment. Of course, the manual alignment is in some sense ``correct'', since expert judgement is used to judge the most biologically-relevant alignment. SA is doing what it is expected to do, in that longer alignments with lower RMSD (after accounting for the gap penalty) will be preferred (the posterior density of such alignments will be higher). DeepAlign incorporates extra information, such as hydrogen bonding, which could explain the better agreement with the reference alignments. Our framework naturally allows for further information such as this to be incorporated.

\subsection{Comparison $3$: RIPC data set}\label{sec:RIPC}

The RIPC data set \citep{Mayr2007} contains examples of particularly challenging test cases. The full data set consists of $40$ pairs of proteins, each of which exhibits one or more of the following challenging features: (R)epetitions, (I)ndels, (P)ermutations, (C)onformational changes. Due to the sequence order constraint of our prior on the alignment, it is not appropriate to consider permutations (which involve a reordering of subunits in a protein), which leaves $28$ pairs for our analysis.  

For $13$ of the $28$ pairs, there are partial reference alignments, manually-curated, of amino acid residues which align, according to experts. They are not ``complete'' alignments, as they only indicate certain residues which match, which in some cases is only a few. Here, we compare SEQ-ALIBI, DeepAlign and Matt. Other methods which have performed well on the RIPC data set include DEDAL \citep{daniluk2011}, PROMATCH \citep{poleksic2016}, PAUL \citep{wohlers2010} and MATRAS \citep{kawabata2003}.  

%\begin{figure}[!h]
%\begin{minipage}{0.45\textwidth}
%\begin{center}
%\includegraphics[angle=270,scale=0.40]{RIPC_RMSDBoxplot.ps}\\
%\end{center}
%\end{minipage}
%\begin{minipage}{0.45\textwidth}
%\begin{center}
%\includegraphics[angle=270,scale=0.40]{RIPC_LALIBoxplot.ps}\\
%\end{center}
%\end{minipage}
%\caption{RMSDs (left) and alignment lengths (right) for SEQ-ALIBI, DeepAlign and MATT.}\label{fig:RIPCRMSD}
%\end{figure}

\begin{figure}[!h]
\begin{minipage}{0.24\textwidth}
\begin{center}
\includegraphics[angle=270,scale=0.4]{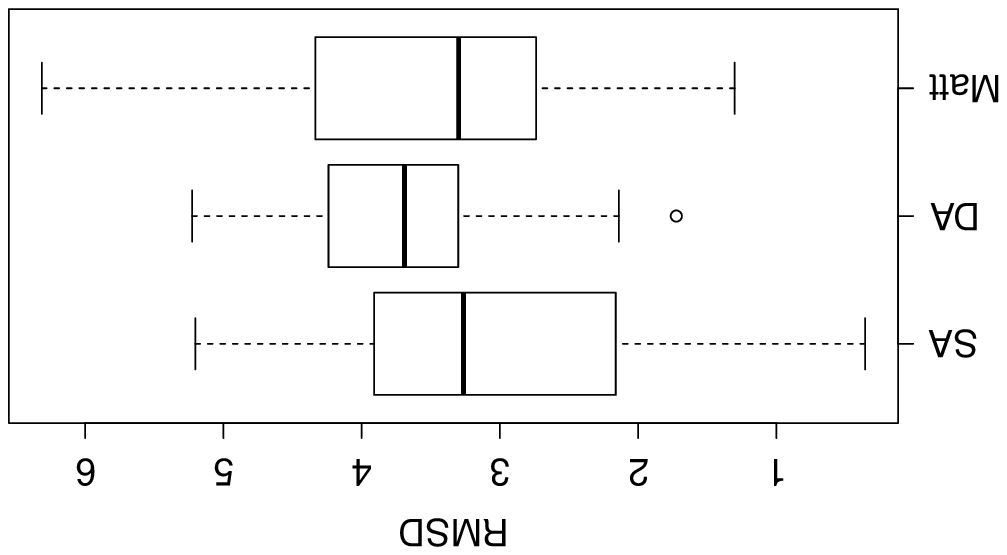}\\
\end{center}
\end{minipage}
\begin{minipage}{0.24\textwidth}
\begin{center}
\includegraphics[angle=270,scale=0.4]{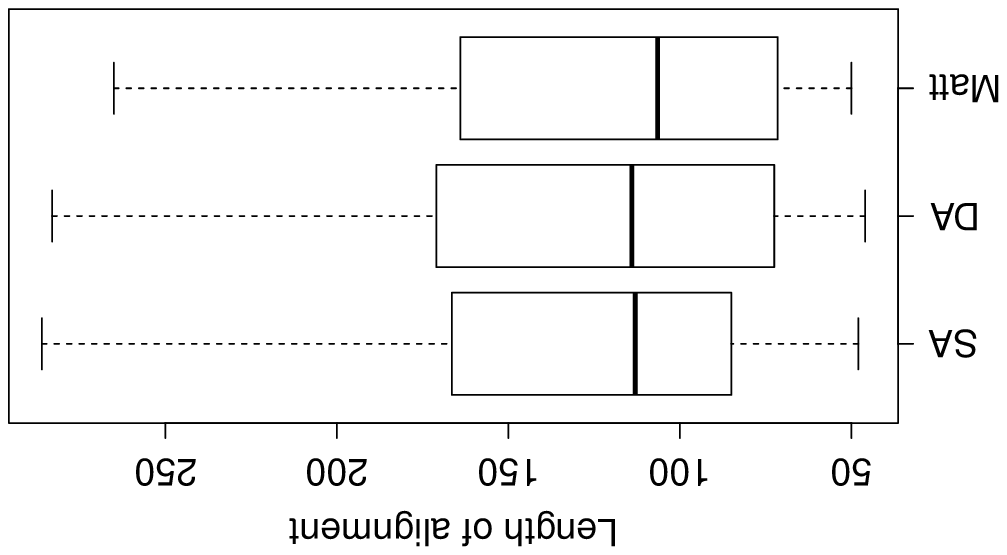}\\
\end{center}
\end{minipage}
\begin{minipage}{0.24\textwidth}
\begin{center}
\includegraphics[angle=270,scale=0.4]{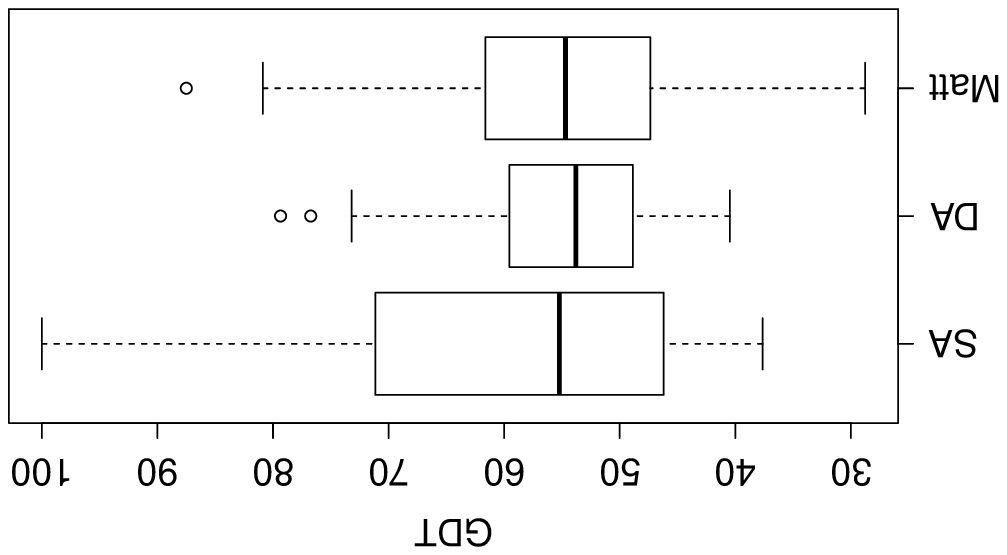}\\
\end{center}
\end{minipage}
\begin{minipage}{0.24\textwidth}
\begin{center}
\includegraphics[angle=270,scale=0.4]{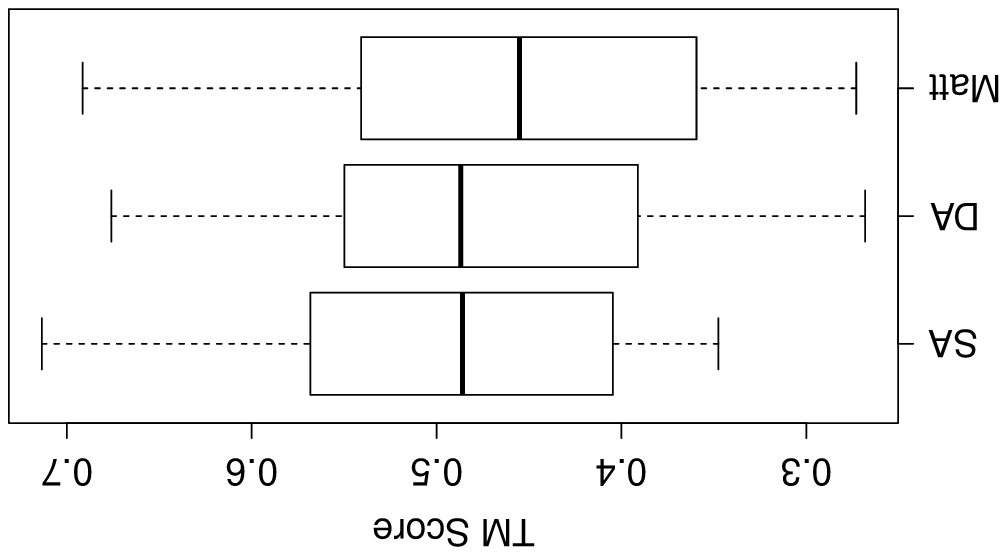}\\
\end{center}
\end{minipage}
\caption{RMSDs, alignment lengths, GDT scores and TM scores, respectively, for SEQ-ALIBI, DeepAlign and Matt on the RIPC data set.}\label{fig:RIPCRMSD}
\end{figure}

The alignments from SEQ-ALIBI have lower RMSD overall (Figure \ref{fig:RIPCRMSD}). The distribution of alignment lengths is similar to that of DeepAlign, with Matt tending to give shorter alignments. SEQ-ALIBI tends to find alignments with higher GDT and TM scores (Figure \ref{fig:RIPCRMSD}).

%\includegraphics[scale=0.57,angle=270,trim={0cm 0.3cm 2cm 0},clip]{GDTBoxplot.ps}

%\begin{figure}[!h]
%\begin{minipage}{0.45\textwidth}
%\begin{center}
%\includegraphics[angle=270,scale=0.25]{RIPC_GDTBoxplot.ps}\\
%\end{center}
%\end{minipage}
%\begin{minipage}{0.45\textwidth}
%\begin{center}
%\includegraphics[angle=270,scale=0.25]{RIPC_TMBoxplot.ps}\\
%\end{center}
%\end{minipage}
%\caption{GDT scores (left) and TM scores (right) for SEQ-ALIBI, DeepAlign and MATT.}\label{fig:RIPCGDTTM}
%\end{figure}

\begin{table}[!h]
\caption{\label{tab:RIPCref} Percentage of the manually-aligned residues found by SEQ-ALIBI, DeepAlign and Matt for RIPC protein data. Also given are the types of challenge presented by each pair (C,I,R) and the number of aligned residues (matches) given in the reference alignment.}
\begin{tabular}{|c|c|c|c|c|c|}\hline
Pair & Challenge & Number of matches  & \% agreement  & \% agreement  & \% agreement  \\ 
 & & in reference alignment &  (SEQ-ALIBI) &   (DeepAlign) &   (Matt) \\\hline 
1  & IC   &   4   &   75   &  75  &   75 \\
2  & I    &  10   &  100   &  90  &  100 \\
3  & I    &   3   &   33   & 100  &  100 \\
4  & I    &   3   &   33   & 100  &  100 \\
5  & IC   &   8   &   75   & 100  &  100 \\
6  & I    &  12   &   67   & 100  &  100 \\
7  & I    &  11   &   91   &  64  &  100 \\
8  & CR   & 148   &   43   &  49  &    0 \\
9  & CR   &   4   &   50   &  50  &   50 \\
10 & IC   &   4   &  100   & 100  &  100 \\
11 &  C   & 101   &   50   &  72  &   85 \\
12 &  C   & 220   &   58   &  70  &   95 \\
13 & IC   &   6   &   67   &  67  &   67 \\\hline
\end{tabular}
\end{table}

Results from the comparison to reference alignments are given in Table \ref{tab:RIPCref}. SEQ-ALIBI does not perform quite as well as DeepAlign and Matt in terms of the percentage agreement with the reference alignment residues. We note that SEQ-ALIBI will prefer solutions with long alignments and lower RMSD, as such solutions will give higher posterior density. The better agreement of DeepAlign and Matt with the reference alignments could possibly be explained by the incorporation of amino acid type and hydrogen bonding information (DeepAlign) and allowing for protein flexibility (Matt). SEQ-ALIBI could be modified to include such information, and we would then expect to see more similarity in the alignments obtained. For example, amino acid type could be readily included by adding an additional factor into the likelihood which quantifies the quality of the primary structure alignment using BLOSUM or PAM matrices, as in \citet{rodriguez14}. DeepAlign incorporates hydrogen bonding by looking at orientation of neighbouring residues and the location of the beta-carbon atoms of each residue. Our likelihood could be modified in a similar fashion to include this additional spatial information; see \citet{mardia07b}, who used beta-carbon information within the framework of the method of \citet{GM} to refine alignments. \citet{schmidler2010} has considered a Bayesian model for flexible shapes using changepoint ideas; with modifications to our model, similar ideas could be incorporated within our framework, to allow for protein flexibility as Matt does.    

It should be noted that since some of the reference alignments involve small numbers of matches, the raw percentage of agreement can exaggerate any discrepancies. For example, for pair 3, there are only 3 reference matches, and the two missed by SEQ-ALIBI can be corrected by inserting a single gap. Hence, SEQ-ALIBI is only one index out from the reference alignment for both these matches. The gap inserted then allows 3 matches between amino acids of the same type, information which was used when curating the reference matches. Therefore, inclusion of amino acid type information in SEQ-ALIBI could balance against the penalty for inserting the required gap, and give alignments more similar to DeepAlign and Matt as suggested above. Similar observations apply to many of the other pairs, and there is less discrepancy between the three methods than it appears from looking at the raw percentage agreement with the reference matches.  

\subsection{Computational complexity}

As MCMC is employed to sample from the joint posterior distribution, the running time is of course relatively slow when judged against the computational methods we compare with, which are specially designed  to quickly find an optimal alignment and run in a matter of seconds. However, they are typically deterministic and return one optimum for a given set of input parameters, assuming the existence of a single global optimum. Multiple runs with different inputs would be needed to try and explore the space of potential  alignments, with no clear principled way for doing this in general. Additionally, it is not clear that multiple ``good'' solutions corresponding to very different alignments (e.g. those requiring different spatial transformations) --- different modes in probability terms --- would be easily identified in this manner \citep{Mayr2007}. Our simulation-based approach has the potential to explore multiple alignments in one run, and identify interesting competing alignments to be followed up, and hence complements these fast and efficient deterministic algorithms. 
 
As an illustration, the example in Section \ref{sec:1GKY2AK3} (configurations of $186$ and $226$ points) runs in around $90$ seconds for a single-chain run, and the parallel tempering implementation using $6$ chains in around $15$ minutes (run on a single core, 3.20GHz CPU). However, we have made no use of the potential for parallelisation, and our implementation is not necessarily the most efficient.  As mentioned above, a major benefit of our approach is a framework which allows the space of alignments to be explored and quantified probabilistically in a fully coherent manner. MCMC is the standard approach for ``exact'' sampling-based inference based on complex probability distributions (i.e. that the target distribution is exactly the one of interest). There is potential for utilising approximations to the true target distribution, which would enable access to quick approximate solutions whilst retaining the fully probabilistic framework.    

%Or incorporate in e.g. practical settings 
%Not optimised, not parallel .... in discussion also, mention of approximate methods (retaining desire for full probability model, and approximate it). 
%Allowing permutations....further work.

\section{Discussion}

In this paper we have presented a fully Bayesian model for the alignment of protein structures. The model is based on that of \citet{GM}, but accounts for the constraint that the sequence ordering of the points in each configuration is meaningful and must be preserved when matching pairs of points, which requires a different prior model for the matching matrix $\bM$. Here we have concentrated on priors built from penalties which are functions of the sequence indices. We have illustrated the potential of this approach using a penalty function which allows the degree of proportionality in the indices defining the alignment to be controlled, which contains a commonly-used gap penalty function as a special case. The formulation allows for other penalty functions to be easily incorporated, and computation will be practical and efficient whenever MCMC updates change only a small number of terms which contribute to the overall penalty.    

\citet{rodriguez14} have also developed a Bayesian model for protein structure alignment; we have used the same prior model for $\bM$ as a starting point, but their method  of sampling alignments from the posterior distribution is quite different to ours, in that an entire new alignment is sampled at each iteration as opposed to the small perturbations of our proposals. Additionally, the authors optimise over the registration parameters, which can be viewed as using a Laplace approximation to the marginal posterior distribution. We have treated the registration parameters as additional unknown parameters about which to draw inference, and sampling them from the posterior allows us to account for the extra uncertainty in the alignment as a result of the uncertainty in these parameters. We note that \citet{kenobi2012} have begun numerical comparisons between the two approaches in a particular situation, namely where rigid-body transformations are used and no sequence order constraint is imposed. The flexibility of the fully Bayesian method to handle different transformations and constraints has been further illustrated in this paper and the papers by \citet{mardia13} and \citet{forbes14}. We have illustrated our method on challenging examples considered previously in the literature, and have shown our method to fare well on benchmark data sets alongside established computational methods, whilst offering more in terms of a fully probabilistic interpretation.

%finds sensible solutions with low RMSD and a high number of matches relative to other methods.
As we have focussed on alignments which preserve sequence order in this paper, our method is not appropriate for situations such as circular permutations of structural subunits \citep{Mayr2007}. Such situations can be handled within our general framework, with appropriate modifications to the model/prior distribution. We note that additional information, such as amino acid types and hydrogen bonding (as in e.g. DeepAlign and Formatt) can be incorporated too. Our focus here has been to show how sequence order-constrained protein structure alignment can be modelled using Bayesian methods for unlabelled shape analysis, and such generalisations will be interesting avenues for further work.   
Though the running time is relatively slow (compared to highly-optimised computational algorithms) due to the use of MCMC (as is standard in Bayesian anlaysis of complex models), our approach offers something different in terms of a principled, fully-probabilistic framework giving access to a full posterior distribution over all possible alignments. In contrast, other methods are rather ad-hoc in nature, do not follow normative inferential principles, and/or deliver conclusions that either do not quantify the uncertainty in alignments, or the joint uncertainty in multiple matches. There is also scope for exploring fast approximations to the target posterior distribution. Our view is that probabilistic methods have their place alongside the more heuristic, but very efficient and successful, computational algorithms, and that both approaches can complement each other in any comprehensive analysis.     
%incorporate other factors, and our framework naturally allows for such factors (and others) to be incorporated, for example through the form of the prior on the alignments, to handle challenging situations such as those in the RIPC data set.

\bibliography{GapPaperRefs}

\begin{thebibliography}{}

\bibitem[Altschul, 1988]{altschul88}
Altschul, S.~F. (1988).
\newblock Generalized affine gap costs for protein sequence alignment.
\newblock {\em Proteins: Structure, Function and Genetics}, 32:88--96.

\bibitem[Altschul et~al., 1990]{altschul90}
Altschul, S.~F., Gish, W., Miller, W., Myers, E.~W., and Lipman, D.~J. (1990).
\newblock Basic local alignment search tool.
\newblock {\em Journal of Molecular Biology}, 215:403--410.

\bibitem[Berman et~al., 2000]{berman00}
Berman, H.~M., Westbrook, J., Feng, Z., Gilliland, G., Bhat, T.~N., Weissig,
  H., Shindyalov, I.~N., and Bourne, N.~E. (2000).
\newblock The protein data bank.
\newblock {\em Nucleic Acids Research}, 28:235--242.

\bibitem[Boomsma et~al., 2008]{Boom_etal08}
Boomsma, W., Mardia, K.~V., Taylor, C.~C., Ferkinghoff-Borg, J., Krogh, A., and
  Hamelryck, T. (2008).
\newblock A generative, probabilistic model of local protein structure.
\newblock {\em Proc Natl Acad Sci U S A}, 105(26):8932--7.

\bibitem[Broderick et~al., 2013]{broderick13}
Broderick, T., Kulis, B., and Jordan, M.~I. (2013).
\newblock {MAD}-{B}ayes: {MAP}-based asymptotic derivations from {B}ayes.
\newblock {\em 30th International Conference on Machine Learning}, 28.

\bibitem[Cheng et~al., 2008]{cheng08}
Cheng, H., Kim, B., and Grishin, N. (2008).
\newblock M{A}{L}{I}{D}{U}{P}: a database of manually constructed structure
  alignments for duplicated domain pairs.
\newblock {\em Proteins: Structure, Function, and Bioinformatics},
  70:1162--1166.

\bibitem[Daniels et~al., 2012a]{Daniels2012a}
Daniels, N.~M., Kumar, A., Cowen, L.~J., and Menke, M. (2012a).
\newblock Touring protein space with {M}att.
\newblock {\em IEEE/ACM Trans. Comput. Biol. Bioinformatics}, 9(1):286--293.

\bibitem[Daniels et~al., 2012b]{Daniels2012}
Daniels, N.~M., Nadimpalli, S., and Cowen, L.~J. (2012b).
\newblock Formatt: Correcting protein multiple structural alignments by
  incorporating sequence alignment.
\newblock {\em BMC Bioinformatics}, 13:259.

\bibitem[Daniluk and Lesyng, 2011]{daniluk2011}
Daniluk, P. and Lesyng, B. (2011).
\newblock A novel method to compare protein structures using local descriptors.
\newblock {\em B{M}{C} Bioinformatics}, 12:344.

\bibitem[Dryden et~al., 2007]{dryden07}
Dryden, I.~L., Hirst, J.~D., and Melville, J.~L. (2007).
\newblock Statistical analysis of unlabeled point sets: comparing molecules in
  chemoinformatics.
\newblock {\em Biometrics}, 63:237--251.

\bibitem[Dryden and Mardia, 2016]{Dryden2016}
Dryden, I.~L. and Mardia, K.~V. (2016).
\newblock {\em Statistical Shape Analysis, 2nd edn}.
\newblock Wiley, Chichester.

\bibitem[Durbin et~al., 1998]{durbin98}
Durbin, R., Eddy, S., Krogh, A., and Mitchison, G. (1998).
\newblock {\em Biological Sequence Analysis: Probabilistic Models of Proteins
  and Nucleic Acids}.
\newblock Cambridge University Press, Cambridge.

\bibitem[Forbes et~al., 2014]{forbes14}
Forbes, P. G.~M., Lauritzen, S., and Moller, J. (2014).
\newblock Fingerprint analysis with marked point processes.
\newblock {\em arXiv - 1407.5809}.

\bibitem[Gerstein and Levitt, 1998]{gerstein98}
Gerstein, M. and Levitt, M. (1998).
\newblock Comprehensive assessment of automatic structural alignment against a
  manual standard, the scop classification of proteins.
\newblock {\em Protein Science}, 7:445--456.

\bibitem[Geyer, 1991]{geyer91}
Geyer, C.~J. (1991).
\newblock Markov chain {M}onte {C}arlo maximum likelihood.
\newblock In {\em Computer Science and Statistics: Proc. 23rd Symp. Interface},
  pages 156--163.

\bibitem[Gibrat et~al., 1997]{gibrat97}
Gibrat, J., Madej, T., Spouge, J., and Bryant, S. (1997).
\newblock The {V}{A}{S}{T} protein structure compar-ison method.
\newblock {\em Biophysical Journal}, 72:298.

\bibitem[Godzik, 1996]{godzik96}
Godzik, A. (1996).
\newblock The structural alignment between two proteins: is there a unique
  answer?
\newblock {\em Protein Science}, 1325--1338.

\bibitem[Golden et~al., 2017]{golden17}
Golden, M., Garcia-Portugues, E., Sorensen, M., Mardia, K.~V., Hamelryck, T.,
  and Hein, J. (2017).
\newblock A generative angular model of protein structure evolution.
\newblock {\em Molecular Biology and Evolution}, 34:2085--2100.

\bibitem[Green, 2015]{green15}
Green, P.~J. (2015).
\newblock {M}{A}{D}-{B}ayes matching and alignment for labelled and unlabelled
  configurations.
\newblock In Dryden, I.~L. and Kent, J.~T., editors, {\em Geometry Driven
  Statistics}, pages 377--389. Wiley, Chichester.

\bibitem[Green and Mardia, 2006]{GM}
Green, P.~J. and Mardia, K.~V. (2006).
\newblock Bayesian alignment using hierarchical models, with applications in
  protein bioinformatics.
\newblock {\em Biometrika}, 93(2):235--254.

\bibitem[Hasegawa and Holm, 2009]{hasegawa09}
Hasegawa, H. and Holm, L. (2009).
\newblock Advances and pitfalls of protein structural alignment.
\newblock {\em Current Opinion in Structural Biology}, 19:341--348.

\bibitem[Herman, 2019]{Herman2019}
Herman, J.~L. (2019).
\newblock Enhancing statistical multiple sequence alignment and tree inference
  using structural information.
\newblock In Sikosek, T., editor, {\em Computational Methods in Protein
  Evolution}, pages 183--214. Springer New York, New York.

\bibitem[Holm and Sander, 1993]{holm93}
Holm, L. and Sander, C. (1993).
\newblock Protein structure comparison by alignment of distance matrices.
\newblock {\em Journal of Molecular Biology}, 233:123--138.

\bibitem[Jonker and Volgenant, 1987]{jonker87}
Jonker, R. and Volgenant, A. (1987).
\newblock A shortest augmenting path algorithm for dense and sparse linear
  assignment problems.
\newblock {\em Computing}, 38:325--340.

\bibitem[Jung and Lee, 2000]{jung2000}
Jung, J. and Lee, B. (2000).
\newblock Protein structure alignment using environmental profiles.
\newblock {\em Protein Engineering}, 13:535--543.

\bibitem[Kawabata, 2003]{kawabata2003}
Kawabata, T. (2003).
\newblock M{A}{T}{R}{A}{S}: A program for 3{D} structure comparison.
\newblock {\em Nucleic Acids Research}, 31:3367--3369.

\bibitem[Kenobi and Dryden, 2012]{kenobi2012}
Kenobi, K. and Dryden, I.~L. (2012).
\newblock Bayesian matching of unlabeled point sets using procrustes and
  configuration models.
\newblock {\em Bayesian Analysis}, 7:547--566.

\bibitem[Kent et~al., 2010]{kent10}
Kent, J.~T., Mardia, K.~V., and Taylor, C.~C. (2010).
\newblock Matching unlabelled configurations and protein bioinformatics.
\newblock Technical report, University of Leeds.

\bibitem[Lennox et~al., 2009]{lennox09}
Lennox, K.~P., Dahl, D.~B., Vannucci, M., and Tsai, J.~W. (2009).
\newblock Density estimation for protein conformation angles using a bivariate
  von {M}ises distribution and {B}ayesian nonparametrics.
\newblock {\em Journal of the American Statistical Association},
  104(486):586--596.
\newblock PMID: 20221312.

\bibitem[Liu and Lawrence, 1999]{liu99}
Liu, J.~S. and Lawrence, C.~E. (1999).
\newblock Bayesian inference on biopolymer models.
\newblock {\em Bioinformatics}, 15:38--52.

\bibitem[Ma and Wang, 2014]{Ma14}
Ma, J. and Wang, S. (2014).
\newblock Algorithms, applications, and challenges of protein structure
  alignment.
\newblock {\em Advances in Protein Chemistry and Structural Biology},
  94:121--175.

\bibitem[Maadooliat et~al., 2016]{maadooliat16}
Maadooliat, M., Zhou, L., Najibi, S.~M., Gao, X., and Huang, J.~Z. (2016).
\newblock Collective estimation of multiple bivariate density functions with
  application to angular-sampling-based protein loop modeling.
\newblock {\em Journal of the American Statistical Association},
  111(513):43--56.

\bibitem[Mardia, 2013]{MardiaRSS13}
Mardia, K.~V. (2013).
\newblock Statistical approaches to three key challenges in protein structural
  bioinformatics.
\newblock {\em Journal of the Royal Statistical Society, Series C},
  62:487--514.

\bibitem[Mardia et~al., 2013]{mardia13}
Mardia, K.~V., Fallaize, C.~J., Barber, S., Jackson, R.~M., and Theobald, D.~L.
  (2013).
\newblock Bayesian alignment of similarity shapes.
\newblock {\em The Annals of Applied Statistics}, 7:989--1009.

\bibitem[Mardia et~al., 2007]{mardia07b}
Mardia, K.~V., Nyirongo, V.~B., Green, P.~J., Gold, N.~D., and Westhead, D.~R.
  (2007).
\newblock Bayesian refinement of protein functional site matching.
\newblock {\em BMC Bioinformatics}, 8:257.

\bibitem[Mayr et~al., 2007]{Mayr2007}
Mayr, G., Domingues, F.~S., and Lackner, P. (2007).
\newblock Comparative analysis of protein structure alignments.
\newblock {\em BMC Structural Biology}, 7:50.

\bibitem[Menke et~al., 2008]{menke08}
Menke, M., Berger, B., and Cowen, L. (2008).
\newblock Matt: Local flexibility aids protein multiple structure alignment.
\newblock {\em PLos Computational Biology}, 4:e10.

\bibitem[Myronenko and Song, 2010]{myronenko10}
Myronenko, A. and Song, X. (2010).
\newblock Point set registration: coherent point drift.
\newblock {\em IEEE Transactions on Pattern Analysis and Machine Intelligence},
  32:2262--2275.

\bibitem[Najibi et~al., 2017]{Najibi17}
Najibi, S.~M., Maadooliat, M., Zhou, L., Huang, J.~Z., and Gao, X. (2017).
\newblock Protein structure classification and loop modeling using multiple
  {R}amachandran distributions.
\newblock {\em Computational and Structural Biotechnology Journal}, 15:243 --
  254.

\bibitem[Ortiz et~al., 2002]{ortiz02}
Ortiz, A.~R., Strauss, C. E.~M., and Olmea, O. (2002).
\newblock {MAMMOTH} (matching molecular models obtained from theory): An
  automated method for model comparison.
\newblock {\em Protein Science}, 11:2606--2621.

\bibitem[Poleksic, 2016]{poleksic2016}
Poleksic, A. (2016).
\newblock Detecting non-trivial protein structure relationships.
\newblock {\em Current Bioinformatics}, 1(2):234--242.

\bibitem[Rangarajan et~al., 1997]{rang97}
Rangarajan, A., Chui, H., and Bookstein, F.~L. (1997).
\newblock The {S}oftassign procrustes matching problem.
\newblock In {\em Information Processing in Medical Imaging}, pages 29--42.
  Springer.

\bibitem[Redelings and Suchard, 2005]{Redlings2005}
Redelings, B.~D. and Suchard, M.~A. (2005).
\newblock {Joint Bayesian Estimation of Alignment and Phylogeny}.
\newblock {\em Systematic Biology}, 54(3):401--418.

\bibitem[Rodriguez and Schmidler, 2014]{rodriguez14}
Rodriguez, A. and Schmidler, S. (2014).
\newblock Bayesian protein structure alignment.
\newblock {\em The Annals of Applied Statistics}, 8:2068--2095.

\bibitem[Schmidler, 2007]{schmidler07}
Schmidler, S.~C. (2007).
\newblock Fast {B}ayesian shape matching using geometric algorithms.
\newblock In Bernardo, J.~M., Bayarri, J., Berger, J.~O., Dawid, A.~P.,
  Heckerman, D., Smith, A.~F., and West, M., editors, {\em Bayesian Statistics
  8}, pages 471--490. Oxford University Press, Oxford.

\bibitem[Schmidler, 2010]{schmidler2010}
Schmidler, S.~C. (2010).
\newblock Bayesian flexible shape matching with applications to structural
  proteomics.
\newblock Technical report, Duke University.

\bibitem[Sela et~al., 2015]{Sela2015}
Sela, I., Ashkenazy, H., Katoh, K., and Pupko, T. (2015).
\newblock {GUIDANCE2: accurate detection of unreliable alignment regions
  accounting for the uncertainty of multiple parameters}.
\newblock {\em Nucleic Acids Research}, 43(W1):W7--W14.

\bibitem[Shih and Hwang, 2004]{shih04}
Shih, E. S.~C. and Hwang, M.~J. (2004).
\newblock Alternative alignments from comparison of protein structures.
\newblock {\em Proteins}, 56:519--527.

\bibitem[Shindyalov and Bourne, 1998]{shindyalov98}
Shindyalov, I.~N. and Bourne, P.~E. (1998).
\newblock Protein structure alignment by incremental combinatorial extension
  ({C}{E}) of the optimal path.
\newblock {\em Protein Engineering design and selection}, 11:739--747.

\bibitem[Srivastava and Jermyn, 2009]{srivastava09}
Srivastava, A. and Jermyn, I.~H. (2009).
\newblock Looking for shapes in two-dimensional cluttered point clouds.
\newblock {\em IEEE Transactions on Pattern Analysis and Machine Intelligence},
  31(9):1616--1629.

\bibitem[Su et~al., 2013]{su13}
Su, J., Srivastava, A., and Huffer, F.~W. (2013).
\newblock Detection, classification and estimation of individual shapes in 2d
  and 3d point clouds.
\newblock {\em Computational Statistics and Data Analysis}, 58:227--241.

\bibitem[Van~Walle et~al., 2005]{Sabmark04}
Van~Walle, I., Lasters, I., and Wyns, L. (2005).
\newblock Sabmark—a benchmark for sequence alignment that covers the entire
  known fold space.
\newblock {\em Bioinformatics}, 21(7):1267--1268.

\bibitem[Wang et~al., 2013]{wang13}
Wang, S., Ma, J., Penf, J., and Xu, J. (2013).
\newblock Protein structure alignment beyond spatial proximity.
\newblock {\em Scientific Reports}, 3:1148.

\bibitem[Wilkinson, 2007]{wilkinson07}
Wilkinson, D.~J. (2007).
\newblock Discussion of ``{F}ast {B}ayesian shape matching using geometric
  algorithms''.
\newblock In Bernardo, J.~M., Bayarri, J., Berger, J.~O., Dawid, A.~P.,
  Heckerman, D., Smith, A.~F., and West, M., editors, {\em Bayesian Statistics
  8}, pages 483--487. Oxford University Press, Oxford.

\bibitem[Wohlers et~al., 2010]{wohlers2010}
Wohlers, I., Domingues, F.~S., and Klau, G.~W. (2010).
\newblock Towards optimal alignment of protein structure distance matrices.
\newblock {\em Bioinformatics}, 26:2273--2280.

\bibitem[Wu et~al., 1998]{wu98}
Wu, T.~D., Schmidler, S.~C., Hastie, T., and Brutlag, D.~L. (1998).
\newblock Regression analysis of multiple protein structures.
\newblock {\em Journal of Computational Biology}, 5:585--595.

\bibitem[Ye and Godzik, 2003]{Ye03}
Ye, Y. and Godzik, A. (2003).
\newblock Flexible structure alignment by chaining aligned fragment pairs
  allowing twists.
\newblock {\em Bioinformatics}, 19 Suppl 2:ii246--55.

\bibitem[Zemla, 2003]{zemla03}
Zemla, A. (2003).
\newblock L{G}{A}: a method for finding 3d similarities in protein structures.
\newblock {\em Nucleic Acids Research}, 31:3370--3374.

\bibitem[Zhang and Skolnick, 2004]{zhang04}
Zhang, Y. and Skolnick, J. (2004).
\newblock Scoring function for automated assessment of protein structure
  template quality.
\newblock {\em Proteins}, 57:702--710.

\bibitem[Zhang and Skolnick, 2005]{zhang05}
Zhang, Y. and Skolnick, J. (2005).
\newblock T{M}-align: A protein structure alignment algorithm based on
  {TM}-score.
\newblock {\em Nucleic Acids Research}, 33:2302--2309.

\bibitem[Zhu et~al., 1998]{zhu98}
Zhu, J., Liu, J.~S., and Lawrence, C.~E. (1998).
\newblock Bayesian adaptive sequence alignment algorithms.
\newblock {\em Bioinformatics}, 14:25--39.

\end{thebibliography}
\bibliographystyle{apalike}

\end{document}